\documentclass[useAMS,usenatbib]{mnras}

\usepackage{newtxtext,newtxmath}

\usepackage[T1]{fontenc}
\usepackage{ae,aecompl}
\usepackage{mn2e-breakabs}

\usepackage{graphicx}	
\usepackage{amsmath}	
\usepackage{booktabs,siunitx} 

\def\gpcoh{\,h^{-1}{\rm Gpc}}
\def\mpcoh{\,h^{-1}{\rm Mpc}}
\def\mpcohcub{\,h^{3}\mathrm{Mpc}^{-3}} 
\def\hMpc{\,\mathrm{Mpc}^{-1}h}

\def\msolaroh{\,h^{-1}M_\odot}

\usepackage[normalem]{ulem}

\title[ELG Mock Challenge]{The Completed SDSS-IV extended Baryon Oscillation Spectroscopic Survey: N-body Mock Challenge for the eBOSS Emission Line Galaxy Sample}

\author[Alam et. al.]{\parbox{\textwidth}{
Shadab Alam \orcid{0000-0002-3757-6359},$^{1}$\thanks{E-mail: salam@roe.ac.uk},
Arnaud de Mattia,$^{2}$
Am\'elie Tamone,$^{3}$
S. \'Avila,$^{4,5}$
John A. Peacock,$^{1}$
V. Gonzalez-Perez,$^{6,7}$,
Alex Smith$^{8}$,
Anand Raichoor$^{3}$ ,
Ashley J. Ross$^{9}$,
Julian E. Bautista$^{10}$,
Etienne Burtin$^{2}$,
Johan Comparat,$^{11}$
Kyle S. Dawson$^{12}$,
H\'elion du Mas des Bourboux$^{12}$,
St\'ephanie Escoffier$^{13}$,
H\'ector Gil-Mar\'in$^{14,15}$,
Salman Habib$^{16,17}$,
Katrin Heitmann$^{16}$,
Jiamin Hou$^{18}$,
Faizan G. Mohammad$^{19,22}$,
Eva-Maria Mueller$^{20}$,
Richard Neveux$^{2}$,
Romain Paviot$^{13,21}$,
Will J. Percival$^{19,22,23}$,
Graziano Rossi$^{24}$,
Vanina Ruhlmann-Kleider$^{2}$,
Rita Tojeiro$^{25}$,
Mariana Vargas Maga\~na$^{26}$,
Cheng Zhao$^{3}$,
Gong-Bo Zhao$^{27,28,10}$,
 } \\ \\
 \scriptsize $^{1}$ Institute for Astronomy, University of Edinburgh, Royal Observatory, Blackford Hill, Edinburgh, EH9 3HJ , UK\vspace*{-2pt} \\
\scriptsize $^{2}$ IRFU,CEA, Universit\'e Paris-Saclay, F-91191 Gif-sur-Yvette, France\vspace*{-2pt} \\  
\scriptsize $^{3}$ Institute of Physics, Laboratory of Astrophysics, Ecole Polytechnique F\'ed\'erale de Lausanne (EPFL), Observatoire de Sauverny, 1290 Versoix, Switzerland\vspace*{-2pt} \\ 
\scriptsize $^{4}$Departamento de F\'isica Te\'orica, Facultad de Ciencias, Universidad Aut\'onoma de Madrid, 28049 Cantoblanco, Madrid, Spain\vspace*{-2pt} \\
\scriptsize $^{5}$Instituto de Fsica Teorica, UAM-CSIC, Universidad Autonoma de Madrid, 28049 Cantoblanco, Madrid, Spain\vspace*{-2pt} \\
\scriptsize $^6$Institute of Cosmology and Gravitation, Portsmouth University, Burnaby Road, Portsmouth PO13FX, UK\vspace*{-2pt} \\
\scriptsize $^7$Astrophysics Research Institute, Liverpool John Moores University, 146 Brownlow Hill, Liverpool L3 5RF, UK\vspace*{-2pt} \\
\scriptsize $^{8}$ IRFU,CEA, Universit\'e Paris-Saclay, F-91191 Gif-sur-Yvette, France\vspace*{-2pt} \\
\scriptsize $^{9}$ Center for Cosmology and AstroParticle Physics, The Ohio State University, Columbus, OH\vspace*{-2pt} \\
\scriptsize $^{10}$ Institute of Cosmology \& Gravitation, Dennis Sciama Building, University of Portsmouth, Portsmouth, PO1 3FX, UK\vspace*{-2pt} \\ 
\scriptsize $^{11}$Max-Planck-Institut f\"{u}r extraterrestrische Physik (MPE), 
Giessenbachstrasse 1, D-85748 Garching bei M\"unchen, Germany\\
\scriptsize $^{12}$
Department of Physics and Astronomy, 
University of Utah, Salt Lake City, UT 84112, USA\vspace*{-2pt} \\
\scriptsize $^{13}$
Aix Marseille Univ, CNRS/IN2P3, CPPM, Marseille, France\vspace*{-2pt} \\
\scriptsize $^{14}$
Institut de Ci\`encies del Cosmos,  
Universitat  de  Barcelona,  ICCUB,  
Mart\'i  i  Franqu\`es  1,  E08028  Barcelona,  Spain\vspace*{-2pt} \\
\scriptsize $^{15}$
Institut  d’Estudis  Espacials  de  Catalunya  (IEEC),  
E08034  Barcelona,  Spain\vspace*{-2pt} \\
\scriptsize $^{16}$ High Energy Physics Division, Argonne National Laboratory, Lemont, IL 60439, USA\vspace*{-2pt} \\ 
\scriptsize $^{17}$ Computational Science Division, Argonne National Laboratory, Lemont, IL 60439, USA\vspace*{-2pt} \\ 
\scriptsize $^{18}$
Max-Planck-Institut f\"ur Extraterrestrische Physik, 
Postfach 1312, Giessenbachstr., 85748 Garching bei M\"unchen, Germany\vspace*{-2pt} \\
\scriptsize $^{19}$
Waterloo Centre for Astrophysics, 
University of Waterloo, Waterloo, ON N2L 3G1, Canada\vspace*{-2pt} \\
\scriptsize $^{20}$ Sub-department of Astrophysics, Department of Physics, University of Oxford, Denys Wilkinson Building, Keble Road, Oxford OX1 3RH\vspace*{-2pt} \\ 
\scriptsize $^{21}$ Aix Marseille Univ, CNRS, CNES, LAM, Marseille, France\vspace*{-2pt} \\
\scriptsize $^{22}$ 
Department of Physics and Astronomy, University of Waterloo,
Waterloo, ON N2L 3G1, Canada\vspace*{-2pt} \\
\scriptsize $^{23}$
Perimeter Institute for Theoretical Physics, 
31 Caroline St. North, Waterloo, ON N2L 2Y5, Canada\vspace*{-2pt} \\
\scriptsize $^{24}$
Department of Physics and Astronomy, 
Sejong University, Seoul, 143-747, Korea\vspace*{-2pt} \\
\scriptsize $^{25}$
School of Physics and Astronomy, 
University of St Andrews, 
St Andrews, KY16 9SS, UK\vspace*{-2pt} \\
\scriptsize $^{26}$
Instituto de F\'isica, 
Universidad Nacional Aut\'onoma de M\'exico, 
Apdo. Postal 20-364, Ciudad de M\'exico, M\'exico\vspace*{-2pt} \\
\scriptsize $^{27}$
National Astronomy Observatories, 
Chinese Academy of Science, 
Beijing, 100012, P. R. China\vspace*{-2pt} \\
\scriptsize $^{28}$
College of Astronomy and Space Sciences, 
University of Chinese Academy of Sciences, Beijing 100049,
China\\
}

\newcommand{\orcid}[1]{\href{https://orcid.org/#1}{\includegraphics[width=0.7em]{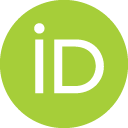}}}

\pubyear{2020}

\begin{document}

\date{Submitted to MNRAS} 

\pagerange{\pageref{firstpage}--\pageref{lastpage}} \pubyear{2020}
\maketitle
\label{firstpage}

\begin{abstract}
Cosmological growth can be measured in the redshift space clustering of galaxies targeted by spectroscopic surveys.
Accurate prediction of clustering of galaxies will require understanding galaxy physics which is a very hard and highly non-linear problem.
Approximate models of redshift space distortion (RSD) take a perturbative approach to solve the evolution of dark matter and galaxies in the universe. 

In this paper we focus on eBOSS emission line galaxies (ELGs) which live in intermediate mass haloes. We create a series of mock catalogues using haloes from the Multidark and {\sc Outer Rim} dark matter only N-body simulations. Our mock catalogues include various effects inspired by baryonic physics such as assembly bias and the characteristics of satellite galaxies kinematics, dynamics and statistics deviating from dark matter particles.

We analyse these mocks using the TNS RSD model in Fourier space and the CLPT in configuration space. We conclude that these two RSD models provide an unbiased measurement of redshift space distortion within the statistical error of our mocks. We obtain the conservative theoretical systematic uncertainty of $3.3\%$,  $1.8\%$ and $1.5\%$ in $f\sigma_8$, $\alpha_{\parallel}$ and $\alpha_{\bot}$ respectively for the TNS and CLPT models. We note that the estimated theoretical systematic error is an order of magnitude smaller than the statistical error of the eBOSS ELG sample and hence are negligible for the purpose of the current eBOSS ELG analysis.
\end{abstract}



\section{Introduction}
\label{sec:intro}

One fundamental consideration in all astronomical studies has remained the same since the begining of astronomy.
That is, the brighter galaxies are more easily detected, up to larger distances, than fainter ones. In the era of large spectroscopic follow-up  \citep{2dFGRS,BOSS,WiggleZ,6dFGRS,Vipers,gama2015,2016AJ....151...44D}, another important metric one has to consider is the ability to  measure the redshift of galaxies. In general this is a function of the line-flux/features in the galaxy spectral energy distribution (SED) widely known as galaxy spectrum.

One of the galaxy population known as Emission Line Galaxies (ELGs) can be Active Galactic Nuclei (AGN) or star forming galaxies. Cosmological surveys are targeting star-forming ELGs for massive spectroscopic surveys \citep{2013MNRAS.428.1498C} at $z \approx 0.5-2$, as: 
\begin{itemize}
    \item There are plenty of ELGs at that epoch when the universe had a higher star-formation density.
    \item  They can provide a spectroscopic redshift measurement  with a short exposure time thanks to strong emission lines, without needing to detect the continuum. 
    \item There emission line can be detected using optical and near-infrared detectors.
\end{itemize}
This has led to the popularity of star-forming ELGs among the architects of galaxy redshift surveys. Hereafter we will use ELGs to refer to star-forming galaxies with strong emission lines. Such characteristics of ELGs has led to focused ELG program in eBOSS survey \citep{2016AJ....151...44D} and also one of the main target sample for ongoing DESI survey \citep{2016arXiv161100036D} which aims to allocate more than 50\% of its fibre budget to ELGs, leading to most precise distance constraint. Other surveys that have or will be targeting ELGs include: 
Euclid~\citep{laureijs11}, PFS~\citep{takada14}, WiggleZ~\citep{WiggleZ}, WFIRST~\citep{spergel2015}, 4MOST~\citep{deJong14}.

Targeting special kind of galaxies means interpreting the cosmological information may become harder due to possibility of complex galaxy formation physics leaking into cosmological measurements.
Star-forming galaxies typically appear blue and generally avoid very high densities ~\citep[e.g][]{chen2017,kraljic2018}.
ELGs are expected to be predominantly lower mass galaxies compared to Luminous red galaxies (LRGs) avoiding the center of massive haloes ~\citep[e.g.][]{favole16,gp18,guo2019}. The complex interplay between the cosmic web and galaxy formation processes makes it more difficult to predict the dark matter halos which hosts such galaxies. eBOSS and DESI aim to target a specific sub-sample of ELGs with high emission in [O$_\textsc{II}$] 3726-3729 \AA\  line flux. Models of galaxy formation show that formation efficiency and dynamics of such galaxy samples are sensitive to the cosmic web beyond the local density and halo mass \citep[e.g.][]{2020MNRAS.491.4294K, gp2020}. Therefore, this sample will have great potential to advance our understanding of galaxy formation physics.

One of the key measurement regarding such galaxy population is the mean host halo mass or linear galaxy bias, which quantify the amplitude of the galaxy clustering compared to the dark matter. \cite{2013MNRAS.433.1146C} studied various photometric selection of ELG samples and measured a galaxy bias being greater than 1.5 using angular clustering and weak lensing. \cite{2013ApJ...767...89M} measured galaxy bias to be 1.3-2.1 of star forming galaxies around redshift 1 and strongly correlated with the star formation rate using the DEEP2 survey \citep{Deep2013}.

The measured redshift of galaxy consists of two components, one is the shift due to expansion of the Universe called cosmological redshift and another is the Doppler shift due to relative velocities projected along the line-of-sight. But in individual galaxy spectrum it is impossible to separate the two. In principle the redshift can also be sensitive to various relativistic effects including gravitational redshift \citep{Cappi1995}. But such effects are very small and negligible for the purpose of this study \citep{2017MNRAS.471.2345Z,2017MNRAS.470.2822A}.
The redshift space clustering of galaxies is not isotropic as distance to the galaxy inferred from their redshift are correlated with their line-of-sight velocity. This produces a distortion in the galaxy correlation function/power spectrum along line-of-sight compared to the plane of sky. This is known as redshift space distortions (RSDs) \citep{Peebles1980,Kaiser87}. The distortion pattern is a measure of galaxy peculiar velocities and hence provides measurement of growth rate at the epoch of the sample called $f$. On very large scales (above $\approx 50\mpcoh$) the clustering of galaxies and their peculiar velocities behave linearly and therefore can be modelled with linear perturbation theory \citep{Kaiser87,Hamilton92}. But most precise measurement of galaxy clustering are obtained at quasi-linear ($\approx 35\mpcoh$) and non-linear ($\approx 10\mpcoh$) scales \citep{Reid14}.
Therefore, it is crucial to be able to model the redshift space clustering measurement at these quasi-linear scales. There have been several recent theoretical developement to extend the linear perturbation theory by performing various expansions and higher order calculations \citep[e.g.][]{2008PhRvD..78h3519M, 2010PhRvD..82f3522T,Carlson12,2014JCAP...05..003O,2016JCAP...12..007V}.  

One of the primary cosmological goals of galaxy redshift surveys is to measure the angular diameter distance ($D_M(z)$), the Hubble constant ($D_H(z)=c/H(z)$) and  the growth rate of structure ($f\sigma_8(z)$) through RSD. Where $\sigma_8$ is the amplitude of the matter fluctuation at $8\mpcoh$ scale. Such measurements when combined with results from the Cosmic Microwave Background \citep{Wmap2013,2018arXiv180706209P} provide the strongest constraints on the ingredients of the Universe such as the amount of dark matter and the geometry of the Universe \citep{2017MNRAS.470.2617A}. This also provides some of the strongest constraints on models of modified gravity, in particular for those driven by measurements of the growth rate ($f\sigma_8$) \citep[e.g.][]{2016MNRAS.456.3743A,2018MNRAS.475.2122M}.

In this paper we focus on two models of RSD namely TNS \citep{2010PhRvD..82f3522T} for the power spectrum and CLPT-GSRSD \citep{Wang13} for the correlation function. Ideally one needs to test the RSD models with mock catalogues produced by solving full physics of galaxy formation along with dark matter dynamics. But currently the best simulation of structure formation known as hyro-dynamical simulations involve various approximation and do not completely reproduce the observed galaxy colour and clustering (see Figure 8 and 16 in \cite{2020arXiv200701889R}). Such hydro-dynamical simulations are also computationally expensive and can only be produced in small volume \citep{2010MNRAS.402.1536S,2015MNRAS.446..521S,dubois2014,2017MNRAS.465.2936M,2018MNRAS.473.4077P,2019MNRAS.486.2827D}. Therefore, we adapt halo occupation distribution (HOD) models \citep[][]{Benson2000,Seljak2000,Peacock2000,White2001,Berlind2002,Cooray2002} for ELG using N-body simulations to produce mock galaxy catalogues occupying large volumes \citep[e.g.][]{2019arXiv191005095A}. We first test the RSD models through a series of non-blind mocks with a variety of baryonic physics added to the mock catalogues. The analysis choices such as priors, range of scales etc., were fixed based on tests on these non-blind mocks. The non-blind mocks means all the true cosmological parameters of the mocks is known to the group analyising  them. We then follow our tests with a set of blind mock in order to avoid any confirmation bias present in the initial non-blind analysis. The blind mocks use known underlying cosmology but has  an unkown value for the growth rate which is revealed only after the analysis is finished. This allowed us to asses the presence of any systematic biases in the measurements, arising from limitations in the theoretical RSD models.  

This study is part of a series of papers analysing the complete eBOSS sample from data release 16 (DR16). Table~\ref{tab:DR16papers} provide a full list of the papers involved in obtaining cosmological constraint from eBOSS DR16.  This paper is organised as follows. We first describe the eBOSS ELG sample in \S ~\ref{sec:data}. 
The models of redshift space distortions are described in \S ~\ref{sec:RSDmodels}.
The N-body simulations used in this paper are described in \S ~\ref{sec:simulation}. 
The details on method to obtain summary statistics from galaxy catalogues is given in \S ~\ref{sec:Measurements}.
The models for ELGs are described in \S ~\ref{sec:ELGmodels}.  The details on unblinded tests of RSD model given in \S ~\ref{sec:nonblind} and blinded tests are discussed in \S ~\ref{sec:blind}. We finally provide the systematic errors in the RSD models in \S ~\ref{sec:sys} and conclude in \S ~\ref{sec:conclusion}.

\begin{table}
	\centering
	\caption{The eBOSS final cosmological interpretation is presented in \citet{eboss20} and galaxy catalogues are described in \citet{anand20,ross20,lyke20a}. Mock catalogues used for covariance matrix and systematic studies is described in \citet{zhao20,lin20a}. A summary of all SDSS BAO and RSD measurements with accompanying legacy figures can be found \href{https://www.sdss.org/science/final-bao-and-rsd-measurements/}{here}.  The full cosmological interpretation of these measurements can be found \href{https://www.sdss.org/science/cosmology-results-from-eboss/}{here}. Analysis for each of the tracers are presented in papers given below.}
	\begin{tabular}{p{1.1cm} p{1.7cm} p{1.7cm} p{2.7cm}} \hline\hline
		 Tracers & $\xi_{\ell}(s)$ & $P_{\ell}(k)$ & Mock Challenge \\
		\hline
		ELG  & \citet{tamone20} & \citet{de-mattia20} & This Work\\
		\hline
		LRG  & \citet{bautista20} & \citet{gil-marin20} & \citet{rossi20}\\
		\hline
		QSO & \citet{hou20} & \citet{neveux20} & \citet{smith20} \\
		\hline
		Ly-$\alpha$ & \multicolumn{3}{c}{\citet{duMasdesBourboux2020} }\\
		\hline
	\end{tabular}
	\label{tab:DR16papers}
\end{table}
\section{eBOSS ELG data}
\label{sec:data}

The extended Baryon Oscillation Spectroscopic (eBOSS) \citep{Dawson2013} project is one of the programmes within the wider 5-year Sloan Digital Sky Survey-IV \citep[SDSS-IV][]{2017AJ....154...28B} using BOSS spectrograph \citep{Smee2013} on the 2.5m Sloan Telescope \citep{Gunn2006}. The eBOSS sample
consists of four different types of tracers, namely Luminous Red
Galaxies \citep[LRG; ][]{2016ApJS..224...34P}; Emission Line Galaxies \citep[ELG; ][]{2017MNRAS.471.3955R}; Quasi-Stellar Objects \citep[QSO; ][]{2015ApJS..221...27M} used as direct tracers of the matter field; and QSOs at higher redshifts (z>2.2), for studies of the Ly$\alpha$ forest \citep{2016A&A...587A..41P}. In this paper we are focusing on testing theoretical models for ELGs, a similar tests have been presented in \citet{smith20} for QSO and \citet{rossi20} for LRGs.

The eBOSS ELGs are selected based on high [O$_\textsc{II}$] flux and are expected to be mostly star
forming galaxies typical of the population at high redshift. An earlier study about ELG selection with the SDSS infrastructure was performed by \cite{2013MNRAS.428.1498C,2013MNRAS.433.1146C} and a pilot survey of ELG testing different target selection algorithms is reported in \cite{2016A&A...592A.121C}. The ELG sample in eBOSS is selected
from intermediate release (DR3/DR5) \citep{2016A&A...585A..50R} of the $grz$-photometry of the Dark Energy Camera (DECam) Legacy survey \cite[DECaLS;][]{2019AJ....157..168D}. The
target selection rules for ELGs in the North Galactic Cap (NGC) and South Galactic Cap (SGC) are slightly
different due to the availability of deeper data in the SGC. 
The ELG selection has two parts, the first of which is to select
star forming galaxies corresponding to the [O$_\textsc{II}$] emission and the second is to preferentially select galaxies around in $0.6<z<1.1$ \citep{2015A&A...575A..40C}.
More details of how these rules were derived and additional considerations are discussed in \cite{2017MNRAS.471.3955R}. The final sample consists of $173,736$ number of ELG galaxies covering a combined area of $730$ square degrees, after veto mask applied, in two different fields (NGC,SGC). The final Large Scale Structure catalogue including systematic weights and observational efficiency is described in \citet{anand20}.

\section{Redshift Space Distortions (RSD) models}
\label{sec:RSDmodels}

In this paper we focus on two models of redshift space distortions namely TNS \citep{2010PhRvD..82f3522T} for the power spectrum and CLPT-GSRSD \citep{Wang13} for the correlation function. We briefly summarise the main ingredients of these models below.

\subsection{TNS model}
\label{sec:TNSmodel}
One of the successful analytical model for the redshift space galaxy power spectrum was proposed by \cite{2010PhRvD..82f3522T} and known as TNS model.
The redshift-space power spectrum in the TNS model is given by:
\begin{equation}
    P_{\rm g}(k,\mu) = P_{\rm TNS}(k) D_{\rm FOG}(k,\mu,\sigma_{\rm v})
    \label{eq:PgTNSDFOG}
\end{equation}
where $k$ is the magnitude of the wavenumber, $\sigma_{\rm v}$ representing the velocity dispersion of satellite galaxies and $\mu$ represents the cosine of the angle from the line-of-sight. The $D_{\rm FOG}$ is the Finger-of-God terms which leads to the suppression of the power spectrum due to the randomness of galaxy peculiar velocities at small scales associated with satellite galaxies. We are using a Lorentzian form $D_{\rm FOG}(k,\mu,\sigma_{\rm v})=(1+0.5(k\mu\sigma_{\rm v})^{2})^{-2}$. The TNS model non-linear power spectrum $P_{\rm TNS}(k)$ is given by:
\begin{equation}
    P_{\rm TNS}(k)= P_{\delta \delta}^{g}(k) + 2f\mu^2 P_{\delta\theta}^{g}(k) + f^2\mu^4 P_{\theta \theta}(k) + C_b(b_1) 
    \label{eq:PTNS}
\end{equation}
where $f$ is the growth rate and $b_1$ is the linear galaxy bias. The galaxy-galaxy ($P_{\delta \delta}^{g}(k)$), galaxy-velocity ($P_{\delta \theta}^{g}(k)$), velocity-velocity ($P_{\theta \theta}(k)$) power spectra and the RSD correction term $C_b$ are calculated using RegPT \citep{2010PhRvD..82f3522T} scheme at 2-loop order. 
Note that the bias terms involved in $P_{\delta \delta}^{g}$, $P_{\delta \theta}^{g}(k)$ and $P_{\theta \theta}(k)$ are calculated following \citet{2009JCAP...08..020M} and \citet{2017MNRAS.466.2242B}.
The linear matter power spectrum, which is the input to the perturbative calculation, is computed at the fiducial cosmology using the Boltzmann code \texttt{CLASS}~\citep{2011JCAP...07..034B}. 

The robustness and precision of this theoretical model is tested in this paper using accurate N-body based mocks with diverse galaxy physics models. This model is used to measure the redshift space distortion signal in the eBOSS ELG power spectrum \citet{de-mattia20}.
We suggest \citet{de-mattia20} for further details about the implementation of this model.

\subsection{CLPT-GSRSD model}
\label{sec:CLPTmodel}
The RSD is essentially the effect caused by the convolution of the line-of-sight component of the velocity field with the spatial distribution of galaxies i.e. the galaxy clustering \citep{ReiWhi11}. Therefore, a simple approach to model the redshift space correlation function of galaxies is by proposing a model for this convolution along with a model to predict the galaxy clustering and velocity field. In this model we used a Gaussian Streaming \citep{ReiWhi11,Wang13} model for the convolution and the Convolution Lagrangian Perturbation Theory \citep[CLPT; ][]{Carlson12} to predict the inherent galaxy clustering and velocity field. 

The redshift space correlation function ($\xi(s_\parallel,s_\bot)$) as the a function of redshift space separations, along line-of-sight ($s_\parallel$) and perpendicular to the line-of-sight ($s_{\bot}$), in the Gaussian streaming model (GSRSD) can be written as follows:
\begin{equation}
1+\xi(s_\parallel,s_\bot) = \int  (1 + \xi(r)) \mathcal{G}(s_\parallel-r_\parallel, v_{12}, \sigma_{12}) d^3r
\end{equation}
where $\xi(r)$, $v_{12}(r)$ and $\sigma_{12(r)}$ are the real space correlation function, the pairwise infall velocity and the pairwise velocity dispersion as the function of real space separation ($r$) between a pair of galaxies. $\mathcal{G}$ describes the probability that a pair of galaxies with separation along the line-of-sight ($r_\parallel$) in real space have a separation ($s_\parallel$) in redshift space.  $\mathcal{G}$ is given by following equation:
\begin{equation}
\mathcal{G}(s_\parallel-r_\parallel, v_{12}, \sigma_{12}) = \frac{1}{\sqrt{2 \pi \sigma_{12}^2(r,\mu)}} \exp \left( \frac{(s_\parallel - r_\parallel - \mu v_{12})^2}{2\sigma_{12}^2(r,\mu)} \right).
\end{equation}
where the real space statistics [$\xi(r)$, $v_{12}(r)$, $\sigma_{12(r)}$] are calculated using the CLPT which is based on the Lagrangian Perturbation Theory \citep[LPT; ][]{Mat08a, Mat08b}. LPT focuses on solving equation of motion of the universe for the displacement field perturbatively as follows:
\begin{equation}
\vec{\psi}(\vec{q},t) = \vec{x}(\vec{q},t)- \vec{q} 
 \approx \vec{\psi}^{(1)} + \vec{\psi}^{(2)} + \vec{\psi}^{(3)} + \vec{\psi}^{(4)}\cdots
\label{eq:disp-field}
\end{equation}

where $\vec{x}(\vec{q},t)$ and $\vec{q}$ are the final and initial positions of the particles at time $t$. The displacement field, $\vec{\psi}(\vec{q},t)$, is expanded as a series of perturbations $\psi^{i}$, where the first order term is the Zel'dovich approximation. The CLPT model identifies terms in the expansion of the density field correlator $<\delta_1\delta_2>$, which become constant in the limit of large scales and kept from being expanded. This essentially leads to a re-summation of LPT with additional terms being exact, leading to a more accurate predictions. Finally, this model takes the linear matter power spectrum, galaxy bias, growth rate and predics the non-linear redshift space correlation function multipoles (see \S~\ref{sec:Measurements} for details) which are used to perform the measurements of mocks N-body galaxy mock catalogue in this paper.
\section{Simulations}
\label{sec:simulation}
In this paper we use dark matter halo catalogues from two different N-body simulations. Mock catalogues are constructed from the simulation snapshot at $z=0.86$, as it is closest in redshift to the effective
redshift of the eBOSS ELG sample ($z_{\rm eff}=0.85$). We briefly describe these simulations in the following subsections.

\subsection{MultiDark Planck 2 (MDPL2)}
\label{sec:sim:MDPL2}
The MultiDark Planck \citep[MDPL2; ][]{2016MNRAS.457.4340K} simulation is publicly available through the CosmoSim
database\footnote{\url{https://www.cosmosim.org/cms/simulations/mdpl2/}} \citep{2012MNRAS.423.3018P,2013AN....334..691R}.
MDPL2 is a N-body simulation run, consists of gravity-only, generated using
the Gadget-2 code.
The simulation assumes a flat $\Lambda$CDM cosmology with $\Omega_m=0.307$, $\Omega_b=0.048$, $h=0.67$, $n_s=0.96$ and $\sigma_8=0.82$. This simulation uses $3840^3$ particles with mass of $1.51\times 10^{9} \msolaroh $ in a periodic box of side length 1000$\mpcoh$. A halo catalogue using the
ROCKSTAR\footnote{\url{https://bitbucket.org/gfcstanford/rockstar}}
phase space halo finder \citep{behroozi13} at an effective redshift of $z \approx 0.86$
snapshot was constructed. ROCKSTAR starts with a
Friends-of-Friends (FoF) group catalogue and analyses particles in full
phase space (i.e. position and velocity) in order to define halo
properties and robustly identify the substructures. 
From the halo catalogue of the simulation, we only use the main halos, removing all the subhaloes and modelling satellite galaxies as described in \S~\ref{sec:ELGmodels}.

\subsection{{\sc Outer Rim}}
\label{sec:sim:outerrim}

The {\sc Outer Rim} N-body simulation (OR) \citep{Habib2016,Heitmann2019} is one of the largest high resolution N-body simulation. {\sc Outer Rim} consists of gravity-only and runs using  Hardware/Hybrid Accelerated Cosmology Code (HACC). 
This simulation uses a flat $\Lambda$CDM WMAP7 \citep{Komatsu2011} cosmology with $\Omega_\mathrm{cdm} h^2=0.1109$, $\Omega_\mathrm{b} h^2=0.02258$, $h=0.71$, $\sigma_8=0.8$ and $n_s=0.963$.
This simulation uses 10{,}240$^3$ particles of mass $m_p=1.85\times 10^9~\msolaroh$ in a periodic box of side length $3~\gpcoh$. 
A halo catalogue using the FoF algorithm \citep{Davis1985} with linking length $b=0.168$ at an effective redshift of $z = 0.865$  snapshot was constructed.

\section{Measurements}
\label{sec:Measurements}

For each of the mock galaxy catalogue, we measure the power spectrum and the correlation function multipoles along with the corresponding covariance matrices. These multipoles are then fitted with the corresponding RSD models in order to perform a measurement of the growth rate along with the geometry of the Universe.

The first step requires obtaining galaxy catalogues with their respective redshift space positions. A given mock galaxy catalogue consists of a list of galaxy positions and velocities in a three-dimensional cubic box with periodic boundary conditions. We first choose one of the axis as line-of-sight. This is then used to determine the redshift space positions of galaxies as follows:
\begin{equation}
    \vec{s}=\vec{r}+\hat{los}\cdot\vec{v}/aH
\end{equation}
where $\vec{s}$, $\vec{r}$ and $\vec{v}$ are the redshift space position, real space position and galaxy velocities in unit of distance. The $\hat{los}$ is the vector pointing to line-of-sight direction, for example if the z-axis is chosen to be the line-of-sight, then $\hat{los}=0\hat{x}+0\hat{y}+1\hat{z}$. Note that the redshift space transformation is performed with periodic boundary conditions. Below we describe the details of this measurement process starting from the galaxy catalogue in redshift space ($\vec{s}$).

\subsection{Measurement in Fourier space [$P_{\ell}(k)$]}

We first take the redshift space galaxy catalogue and estimate the density  contrast ($\delta_{g}(\vec{s})$) on a regular grid of mesh size $512^{3}$ using the Triangular Shaped Cloud (TSC) scheme. The Fourier transformation of the density contrast $\delta_g(\vec{k})$ is then used to estimate the power spectrum as follows:

\begin{equation}
P_{\ell}(k) = \left(2\ell+1\right) \int \frac{d\Omega_{k}}{4\pi V} \delta_{g}(\vec{k})\delta_{g}^{*}(\vec{k})\mathcal{L}_{\ell}(\hat{\vec{k}} \cdot \hat{los}) - P_{\ell}^{\mathrm{noise}}(k) \, .
\label{eq:power_spectrum_estimator_box}
\end{equation}

$\mathcal{L}_{\ell}$ is the Legendre polynomial of order $\ell$ and $\hat{los}$ the chosen line-of-sight vector. $P_{\ell}^{\mathrm{noise}}(k)$ is the shot noise term which is given by the inverse of mean number density for the monopole ($\ell=0$) and is zero otherwise.
The \texttt{nbodykit}~\citep{2018AJ....156..160H} package is used to perform the calculation of the power spectrum. The use of a regular grid to perform the Fast Fourier Transformation (FFT) makes the angular modes distribution irregular at large scales. This affect the final measured power spectrum which we account for in the model by weighting the modes according to the $(k,\hat{\vec{k}} \cdot \hat{los})$ sampling.

We finally fit the TNS model for redshift space power spectrum (see \S~\ref{sec:TNSmodel}) to the measured power spectrum multipoles from N-body simulations. The fit involves three cosmological parameters, which are the growth rate $f=\frac{d \ln D(a)}{d \ln a}$ and two scaling parameters $\alpha_{\bot}= \frac{D_M(z) r^{\rm fid}_s}{D_M^{\rm fid}(z) r_s}$ and $\alpha_{\parallel}=\frac{H^{\rm fid}(z)r^{\rm fid}_s}{H(z)r_s}$. 

\begin{align}
    \alpha_{\bot}= \frac{D_M(z) r^{\rm fid}_s}{D_M^{\rm fid}(z) r_s} \\
    \alpha_{\parallel}=\frac{H^{\rm fid}(z)r^{\rm fid}_s}{H(z)r_s}.
\end{align}
Where $r_s$ is the comoving sound horizon scale. The Alcock-Paczynski
(AP) parameters ($\alpha_{\bot},\alpha_{\parallel}$) compress cosmological information efficiently by re-scaling the distances over the line of sight and perpendicularly to it. Given that the growth rate is degenerate with the normalisation of the power spectrum $\sigma_8$, we always quote measurement of $f\sigma_8$ rather than $f$ itself. Apart from these we also have 4 nuisance parameters, the two bias parameters $b_1,b_2$, one velocity dispersion to account for non-linear Fingers-of-God $\sigma_{v}$ and the stochastic shot noise term $A_g$. The fitted $k$-range of the RSD measurement is $0.02 - 0.2 \hMpc$ for the monopole and quadrupole and $0.02 - 0.15 \hMpc$ for the hexadecapole (see \citet{de-mattia20} for details). We perform a $\chi^{2}$ minimisation using the \texttt{Minuit}~\citep{Minuit1975}\footnote{ {\url{https://github.com/iminuit/iminuit}}} package, with wide priors for all parameters. We perform several tests, including a test on the parameter boundaries to make sure the results are robust. 
Errors on the parameters are given by likelihood profiling at the $\Delta \chi^{2} = 1$ level.

MultiDark mocks are analysed within the fiducial cosmology of eBOSS analyses \citep{de-mattia20} and thus treated as non-periodic; the induced window function effect and global integral constraint are accounted for in the model, following \citet{2017MNRAS.464.3121W}, and \citet{2019JCAP...08..036D}, respectively. The covariance matrix is estimated from $500$ lognormal mocks generated with the MultiDark cosmology, with a bias of $1.4$, and the same density as MultiDark mocks: $3 \times 10^{-3} \; \mpcohcub$.
For the {\sc Outer Rim} mocks we use a Gaussian covariance matrix for the measured power spectrum following the method described in \cite{2016MNRAS.457.1577G}, which has been shown to be accurate enough in the quasi-linear regime probed by RSD analyses.

\subsection{Measurement in Configuration space [$\xi_{\ell}(s)$] }
For the measurement of the galaxy two point correlation function we first perform a pair count of galaxies (called $DD$) in redshift-space as the function of the distance between a pair of galaxies ($s$) and the cosine of the angle of the separation vector from the line-of-sight direction ($\mu$). We then estimate analytically, the pair count (called $RR$) for points  that are uniformly randomly distributed inside the simulation box with the same density as galaxies, using the following equation: 

\begin{equation}
RR(s,\mu)= \frac{N_{\rm gal}(N_{\rm gal}-1)}{L^3_{\rm box}} \left[ \frac{4\pi(s_2^3-s_1^3)}{3} \right] \left[\mu_2-\mu_1 \right] \, .
    \label{eq:RR-count}
\end{equation}
Where $N_{\rm gal}$ and $L_{\rm box}$ are the number of galaxies and the size of the simulation box respectively.  $s_1$ and $s_2$ correspond to the lower and upper limits of the radial bins, while $\mu_1=\cos(\theta_1)$ and  $\mu_2=\cos(\theta_2)$ correspond to the lower and upper limits of the angular bins. We finally obtain the correlation function multipole as follows:
\begin{align}
    \xi_{2D}(s,\mu) &= \frac{DD(s,\mu)}{RR(s,\mu)}-1 \, ,\\
    \xi_{\ell}(s) &= \frac{2\ell+1}{2}\int_{\mu} \xi_{2D}(s,\mu)  \mathcal{L}_{\ell}(\mu) d\mu \, ,
\end{align}
where $\mathcal{L}_{\ell}$ is the Legendre polynomial of order $\ell$. The periodic boundary condition allows the use of an analytic RR pair-count which makes the computation of the correlation function very efficient.

Similar to the power spectrum analysis, we use a Gaussian covariance matrix for the measured correlation function following the method described in \cite{2016MNRAS.457.1577G}.

We finally fit the CLPT model for the redshift space correlation function (see \S~\ref{sec:CLPTmodel}) to the measured correlation function multipoles from N-body simulations. The fit involves three cosmological parameters, which are the growth rate, $f$, and the two scaling parameters, $\alpha_{\bot}$ and $\alpha_{\parallel}$. Similar to the power spectrum analysis, we always quote measurements of $f\sigma_8$ rather than $f$ itself. Apart from these we also have three nuisance parameters, the two Lagrangian bias parameters  $F_1$ and $F_2$, and one velocity dispersion to account for non-linear Fingers-of-God $\sigma_{\rm FOG}$. Only the first order Lagrangian bias ($F_1$) is allowed to be free  and the second order Lagrangian bias ($F_2$) is determined via the peak-background split relation \citep{White14}. The fitted $s$-range of the RSD measurement is $32 - 160 \mpcoh$ for the monopole , the quadrupole and the hexadecapole. We perform a $\chi^{2}$ minimisation using the \texttt{Minuit} package, with wide priors for all parameters.  
We perform several tests similar to the power spectrum analysis to make sure the results are robust. 
Errors on the parameters are given by likelihood profiling at the $\Delta \chi^{2} = 1$ level

\section{Emission Line Galaxies (ELG) models using HOD}
\label{sec:ELGmodels}
Modelling large cosmological volumes of the Universe requires a certain knowledge of galaxy formation. What makes it possible for galaxies to form and what decides properties of these galaxies. The standard model within the hierarchical structure formation suggests that the dark matter collapsing under gravity throughout the evolution of the Universe forms the back bone structure and leads to the formation of the cosmic web \citep{2010gfe..book.....M,2018ARA&A..56..435W}. This cosmic web consists of collapsed dark matter objects called dark matter haloes that are the natural places for galaxies to form. Therefore, the two main popular models to populate large dark matter (N-body) simulations are the Halo Occupation Distribution \citep[HOD;][]{Benson2000,Seljak2000,Peacock2000,White2001,Berlind2002,Cooray2002} and Subhalo Abundance Matching \citep[SHAM:][]{2004ApJ...609...35K,2004ApJ...614..533T,2004MNRAS.353..189V}. These two modelling techniques assume that all galaxies are formed in dark matter halos and that the properties of galaxies are dominantly determined by the mass of the haloes. Alternatively one could use full hydro-dynamical simulations \citep{2010MNRAS.402.1536S,2015MNRAS.446..521S,dubois2014,2017MNRAS.465.2936M,2018MNRAS.473.4077P,2019MNRAS.486.2827D} or Semi Analytical Models \citep[SAMs:][]{2011MNRAS.413..101G,2014MNRAS.439..264G}. The HOD is one of the fastest and simplest way to create mock galaxy catalogues and thus is adequate for the large exploration of different mock catalogues that is done here. 

In the HOD framework we consider two kinds of galaxies in each halo known as the central and satellite galaxies. The occupation recipe provides the probability of a given halo to have a central galaxy and a number of satellite galaxies. There are various degrees of freedom in terms of how the velocities and positions of satellite galaxies are assigned within haloes and they depend on the details of the galaxy population \citep{Reid14, AlamGama2020}. We aim to study a wide variety of HOD models covering a range of physical processes to estimate the robustness of our measurement independently of the details of galaxy physics. In this paper we are using three different parametrisations of the average HOD and a variety of satellite models. Below we describe the three parametrisations used for the shapes of the HOD for central and satellite galaxies.

\subsection{Standard HOD model (SHOD)}
The idea of hierarchical clustering brings a very simple assertion that dark matter haloes with more mass will have more baryons and hence will host more massive galaxies which will also be brighter. 
Therefore, we can simply rank order the dark matter haloes by their mass and galaxies by their brightness and connect them one-to-one with some dispersion. This intuitive picture about the connection between dark matter haloes and galaxies has been remarkably useful. The popular 5 parameter Standard HOD model (hereafter SHOD) is 
shown to describe the mean occupation probability for the detailed hydro-dynamical models and semi-analytical models of galaxy formation \citep{zheng2005,White2011}.

This essentially says that the massive dark matter haloes host galaxies with constant probability and depending on the brightness limit the probability of hosting central galaxy will have a cut-off halo mass. More formally the central occupation probability in this model is parameterised as follows:

\begin{equation}
    p_{\rm cen}^{\rm SHOD}=\left< N_\mathrm{cen}^\mathrm{SHOD} (M_{\rm h}) \right> = \frac{1}{2}p_{\rm max} \mathrm{erfc}\left(\frac{\log_{e}{ M_{\rm c}-\log_{e}{M_{\rm h}}}}{\sqrt{2} \sigma_M}\right) \, .
    \label{eq:Nerrf}
\end{equation}
where $p_{\rm max}$ decides the saturation occupation probability in the high halo mass limit, $M_{\rm c}$ and $\sigma_M$ decides the cut-off halo mass and its dispersion for the given galaxy sample. Models of galaxy formation and evolution have shown that this HOD model is not adequate for star-forming galaxies in general, including star-forming ELGs~\citep[e.g.][]{geach12,contreras13,cochrane2018,gp18} . However the physical processes involved in the formation and growth of ELGs are complex and require more flexibility such as quenching at the centre of massive haloes.

\subsection{High Mass Quenched model (HMQ)}

eBOSS ELGs are expected to avoid residing in the centre of massive halos \citep[e.g.][]{favole16,gp18,guo2019}. Such behaviour is not possible to accommodate in the SHOD model. Therefore, \cite{2019arXiv191005095A} proposed a modified HOD framework encapsulating such behaviour called High Mass Quenched model (HMQ). The occupation probability of central galaxy of a halo is given by the following equation:
\begin{align}
    p_{\rm cen}^{\rm HMQ}=\left< N_\mathrm{cen}^\mathrm{HMQ} (M_{\rm h}) \right> &=  2 A \phi(M_{\rm h}) \Phi(\gamma M_{\rm h})  + & \nonumber \\  
    \frac{1}{2Q} & \left[1+\mathrm{erf}\left(\frac{\log_{e}{M_h}-\log_{e}{ M_c}}{0.01}\right) \right],  \label{eq:NHMQ}\\
\phi(x) &=\mathcal{N}(\log_{e} M_c, \sigma_M), \label{eq:NHMQ-phi}\\
\Phi(x) &= \int_{-\infty}^x \phi(t) dt = \frac{1}{2} \left[ 1+\mathrm{erf} \left(\frac{x}{\sqrt{2}} \right) \right], \label{eq:NHMQ-Phi}\\
A &=\frac{ p_{\rm max}  -1/Q}{ \mathrm{max}_{\rm x}(2\phi(x)\Phi(\gamma x))} \, .
\label{eq:NHMQ-A}
\end{align}
The effect of various parameters on the HMQ occupation function is illustrated in Figure 1 of \cite{2019arXiv191005095A}.The parameter $M_c$ is the cut-off mass of ELG centrals impacting the location of the peak in occupation probability. $Q$ sets the quenching efficiency for high mass haloes; a larger value of $Q$ implies more efficient quenching. The function $\phi(M_h)$ is the normal distribution given in equation~\ref{eq:NHMQ-phi} and $\Phi(M_h)$ is the cumulative density function of $\phi(M_h)$ given in equation~\ref{eq:NHMQ-Phi}. These two functions depend on the parameters $\gamma$ controlling the skewness and $\sigma_M$ controlling the width. The parameter $A$ sets the overall formation efficiency of ELGs given in equation~\ref{eq:NHMQ-A} and depends on $p_{\rm max}$. 

\subsection{Star forming HOD (SFHOD)}

Another way to parametrise the mean HOD of ELGs is based on the results from the semi-analytical model of galaxy formation and evolution presented in~\citet{gp18}, which included a simple approach to model the nebular emission in star-forming galaxies. We call this alternate parametrisation (SFHOD) which was first proposed in \citep{avila20} given below:

\begin{equation}
p_{\rm cen}^{\rm SFHOD} =\langle N_{\mathrm{cen}}^{\mathrm{SFHOD} }(M_h) \rangle =
   \begin{cases} 
      \frac{A_c}{\sqrt{2\pi}\sigma} \cdot e^{-\frac{({\rm log_{10}}M_h-\mu)^2}{2\sigma^2}}  & M_h\leq 10^{\mu} \\
      \frac{A_c}{\sqrt{2\pi}\sigma} \cdot \Bigg( \frac{M_h}{10^\mu}\Bigg)^\gamma & M_h\geq  10^{\mu} \\
   \end{cases}
\label{eq:HMQb}
\end{equation}
The parameter $\mu$ is the logarithm of the halo mass with the highest occupation probability for ELG centrals with $\sigma$ giving its width and $A_c$ overall normalisation. The parameter $\gamma$ suppresses the occupation probability at the high mass ends.

The HMQ and SFHOD functional forms are closer representation of ELGs as per current understanding and expected to produce more realistic host halo distribution as observed in data. Note that the HMQ and SFHOD models will have quite different contribution to non-linearity compared to the SHOD, due to different kind of haloes hosting ELGs in the extreme ends of halo mass distribution.

\subsection{Satellite galaxies}
The number of satellite galaxies as a function of halo mass is given by the following functional form:
\begin{equation}
    p_{\rm sat}\left< N_{\rm sat} (M_{\rm halo}) \right> = A_s \left( \frac{M_{\rm h} - \kappa M_c}{M_1}\right)^\alpha .
\end{equation}
The number of satellite galaxies is assumed to be a power
law with index $\alpha$ and characteristic satellite mass $M_1$. The cut-off mass is set by the parameter $\kappa$ in units of $M_c$ below which the probability of finding a satellite galaxy is zero. The parameter $A_s$ is used to calibrate the amplitude of the satellite occupation. We use the same functional form to model the mean number of satellites for all three models (i.e. SHOD, HMQ, SFHOD), with independent parameters in each case. Satellite galaxies follow a Poisson distribution for the SHOD and HMQ models but it has an additional free parameter $\beta$ for the SFHOD model. In the SFHOD model, $\beta=0$ is equivalent to a Poisson distribution, $0<\beta<1$ correspond to a negative binomial distribution with $p=\frac{1}{1+\beta^2}$. The SFHOD model also allows satellite distribution with Next Integer distribution  as given in equation 22 of \cite{avila20} and labelled as NI.

\begin{figure}
    \centering
    \includegraphics[width=0.49\textwidth]{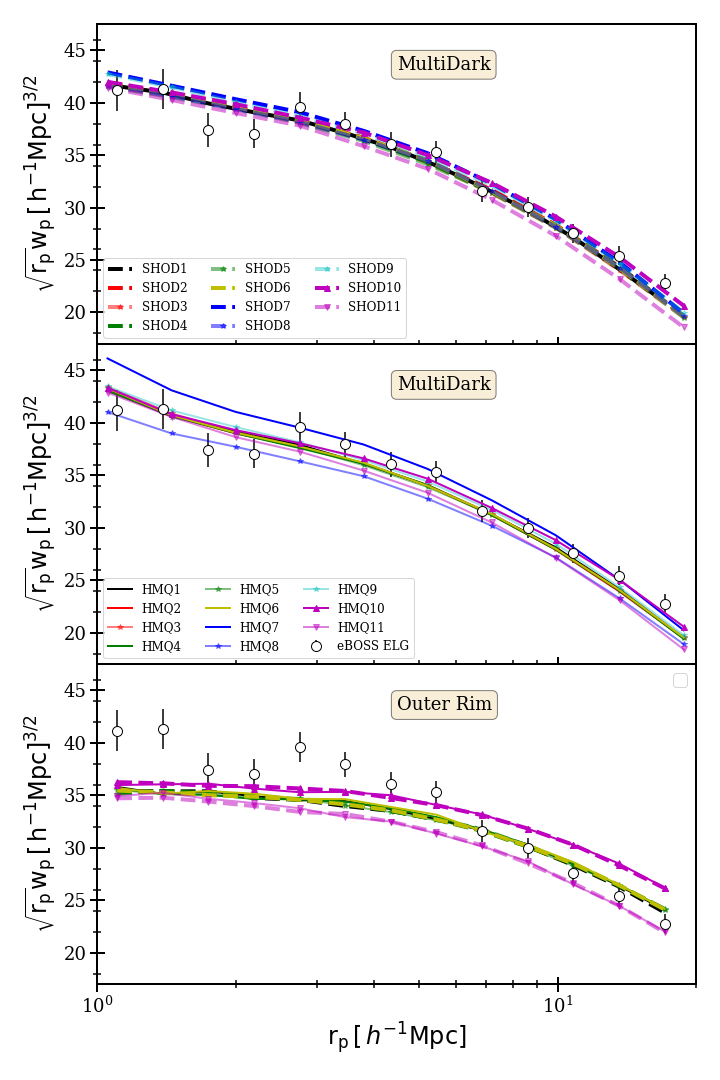}
    \caption{Project correlation function of eBOSS ELG sample along with the various non-blind mocks. The black open circles in all panels represents the $w_p$ measured from eBOSS data. The top and middle panels shows $w_p$ for MultiDark based mocks and the bottom panel is for {\sc Outer Rim} based mocks. The solid coloured lines are for HMQ model and dashed coloured line is for SHOD models. The different colours indicating different types of models as detailed in Table~\ref{tab:HODmodels}.}
    \label{fig:wp-mock-data}
\end{figure}

\section{Mock Challenge}
\label{sec:nonblind}

We create a series of mocks, from a total of $~40$ different models, with variations in the parameters discussed in \S~\ref{sec:ELGmodels} and beyond~\citep{2019arXiv191005095A,avila20}. 
As we discussed in \S~\ref{sec:ELGmodels}, ELG host dark matter haloes properties are still under investigation. Therefore, the main focus here is to explore as many ways as possible to populate dark matter haloes with star forming galaxies, to make sure the real properties of ELGs are encapsulated within the series of mocks we produce. The galaxy mocks created in this section are analysed with known expected parameters, which is a non-blind test of models. The next section describes similar tests for blind mocks. 

The probability of occupying a central galaxy $\left< N_{\mathrm{cen}}\right>$ is evaluated to create a mock galaxy catalogue for each dark matter N-body simulation box. The $\left< N_{\mathrm{cen}}\right>$ is mainly a function of the halo mass but it may depend on other halo properties depending on the details of the model used. We then generate uniform random numbers and populate a central galaxy at the center of the halo with the halo velocity if the random number is below $\left< N_{\mathrm{cen}}\right>$. We then evaluate the mean number of satellite galaxies using $\left< N_{\mathrm{sat}}\right>$ for each halo which again mainly depend on the halo mass but may depend on other halo properties. The actual number of satellites assigned to each halo is generally sampled from a Poisson distribution but for some models it follows different statistics (see \S~\ref{sec:ELGmodels}). Different schemes are assumed by the models to assign the positions of satellite galaxies. They may follow a NFW distribution, a scaled NFW distribution or the distribution of randomly sampled dark matter particles from the halo. The velocities of the satellite galaxies are sampled from the velocity dispersion of halos but some models scale the velocity dispersion by a free parameter to make the satellites hotter or cooler than dark matter particles. Some of the models also introduce an infalling velocity to the satellite. Below we describe the details of the mock catalogues created using two sets of simulation.

\subsection{MultiDark Mocks}
\label{sec:MDPL2-nonblindmocks}

\begin{figure*}
    \centering
    \includegraphics[width=1.0\textwidth]{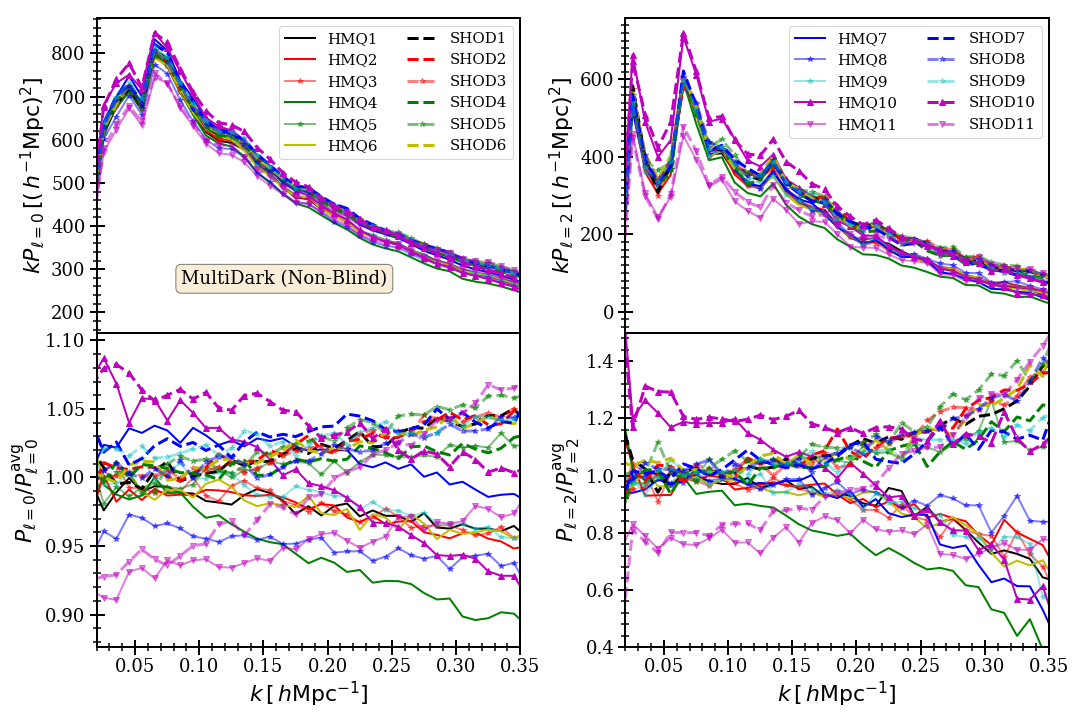}
    \caption{Power spectrum multipoles for mocks from the MultiDark simulation. The top left (right) panel shows the monopole (quadrupole) moment of the power spectrum. The bottom left (right) panel shows the ratio of monopole (quadruple) to the average power spectrum. The solid lines are for the mocks using the HMQ model and dashed are for the SHOD model. The different colors correspond to different models as described in Table~\ref{tab:HODmodels}.}
    \label{fig:MDPL2-nonblind-pk}
\end{figure*}

Best fit HOD parameters for both the SHOD and the HMQ models are obtained by fitting to the measured projected correlation function, the halo catalogue from the MultiDark simulation snapshot at redshift 0.86. The best fit model parameters are obtained by minimising the $\chi^2$ as given below: 

\begin{equation}
    \chi^2=\left[ w_p^{\rm eBOSS}-w_p^{model}\right]^{\rm T} \mathcal{C}^{-1} \left[ w_p^{\rm eBOSS}-w_p^{model}\right]
    \label{eq:chi2}
\end{equation}
where $w_p^{\rm eBOSS}$ is the measured projected correlation function from eBOSS ELG sample and $\mathcal{C}^{-1}$ is the inverse covariance matrix obtained using the jackknife re-sampling scheme following \cite{2019arXiv191005095A}. This fit was performed only for the fiducial HOD model (model=1) and the best fit parameters are given in Table~\ref{tab:HODparameters}. For other variants of the HMQ and SHOD models, we keep these basic parameters fixed allowing the variation in  other degrees of freedom. In principle, one could re-fit the basic HOD parameters along with each of the additional free parameters, but given the large errors in the HOD parameters and our focus on generating a variety with approximately the linear bias of eBOSS ELG sample, we did not performed such a refit. 

The projected correlation function measured from eBOSS data and various MultiDark based mock catalogues are shown in the upper panel of Figure~\ref{fig:wp-mock-data}. The black empty points show the measurement from the eBOSS sample with jackknife errors. The various coloured lines represent the projected correlation function measured from the mock catalogues. Solid lines are for HMQ models and the dashed lines are for SHOD models. The details of the different models are given in Table~\ref{tab:HODmodels}. Note that the differences in the projected clustering for different models mainly stems from the fact that we did not try to refit each model. We also show the redshift space power spectrum multipoles for each of the model in Figure~\ref{fig:MDPL2-nonblind-pk} with the same lines and colours convention as in the $w_p$ plot. The monopole and quadrupole moments are shown in the top left and right panel of the plot. Bottom panels shows the ratio of monopole and qudrupole with respect to the average monopole and quadrupole of all the models. 

Table~\ref{tab:HODmodels} lists the 11 models generated for this paper. Model number 1 (i.e. HMQ1,SHOD1) is the fiducial one with halo mass only HOD and for which haloes are populated with satellites using a NFW profile for the spatial distribution along with dark matter halo velocity dispersion for redshift space positions. The next two models, 2 (i.e. HMQ2,SHOD2) \& 3 (i.e. HMQ3,SHOD3),  modify the concentration by populating satellite galaxies more or less concentrated by a factor of 50\% respectively. The velocity dispersion for satellite galaxies is higher or lower by a factor of 50\% for models number 4 and 5 compared to the velocity dispersion of particles in the halo. 
In model number 6, we allow central galaxies to be shifted from the centre of the dark matter halo following a Gaussian distribution with width of $0.1r_{200}$. Models number 7, 8 and 9 have assembly bias by setting the occupation of central, satellites and both (i.e. central and satellites) 
to be correlated with the dark matter haloes concentration parameter. We follow the scheme suggested in \cite{zentner2016} for models with assembly bias where the occupation numbers of central and satellite galaxies are modified using the following equation:

\begin{equation}
    \langle N_{\rm cen,sat} \rangle(M_h,c)= \langle N_{\rm cen,sat} \rangle(M_h) + (-1)^{p(c)} \delta N_{\rm cen,sat}(M_h,c)
    \label{eq:assembly1}
\end{equation}
where $\langle N_{\rm cen,sat} \rangle(M_h)$ is the  standard occupation number of central or satellite galaxies as detailed in \S~\ref{sec:ELGmodels}. The parameter $c$ represents the dark matter halo concentration parameter. The functions $\delta N_{\rm cen}(M_h,c)$ and $\delta N_{\rm sat}(M_h,c)$ are given by following equations:
\begin{align}
    \delta N_{\rm cen} &= A_{\rm cen} \, \mathcal{MIN} \left[ \langle N_{\rm cen} \rangle(M_h),1-\langle N_{\rm cen} \rangle(M_h) \right] \\
    \delta N_{\rm sat} &= A_{\rm sat}\langle N_{\rm sat} \rangle(M_h)
    \label{eq:deltaN}
\end{align}
where $A_{\rm cen}$ and $A_{\rm sat}$ are the two free parameters which control the level of assembly bias. The function $p(c)$ is a step function with $p(c)=0$ for $c>= c_{\rm median}$ and $p(c)=1$ for $c<c_{\rm median}$, where $c_{\rm median}$ is the median concentration of all the dark matter haloes. Models number 10 and 11 have higher or lower peculiar velocities by a factor of 20\%, which allows the growth rate ($f$) of the constructed mock catalogues to be altered keeping fixed all other parameters.

\begin{table}
	\centering
	\caption{The best fit parameters for different HOD models and different N-body simulations. The first two columns corresponds to MultiDark and the next two for {\sc Outer Rim}. Note that the details of the SFHOD models are given in Table~\ref{tab:SFHODmodel}.}
	\begin{tabular}{l|cc|cc} 
		\hline
		\multicolumn{1}{l}{Parameters} & \multicolumn{2}{c}{MultiDark} & \multicolumn{2}{c}{{\sc Outer Rim}} \\
		                  & SHOD  & HMQ\         & SHOD & HMQ\\
		\hline
		$\log_{10}(M_c)$  & 11.70  & 11.6 & 11.4 & 11.5 \\
		$\sigma_M$        & 0.59  & 0.61  & 0.1  & 0.61 \\
		$\gamma$          & -     & 4.04  & -    &  4.04   \\
		$Q$               & -     & 100   & -    & 100  \\
		$\log_{10}(M_1)$  & 14.4  & 13.55 & 13.6 & 13.55  \\
		$\kappa$          & 1.0   & 1.0     & 1.0  & 1.0  \\
		$\alpha$          & 0.40  & 0.99  & 0.4  & 0.99  \\
		\hline
	\end{tabular}
	\label{tab:HODparameters}
\end{table}

\begin{table*}
    \centering
    \caption{ List of SHOD and HMQ HOD models with their detailed description and simulations used. The basic HOD parameters used for these models are given in Table~\ref{tab:HODparameters} with any additional degree of freedom described in this table.}
    \begin{tabular}{p{1.2cm} p{11cm} p{2.0cm}} \hline\hline
    {\bf Model} &  {\bf Description} & {\bf Simulations} \\ [0.5ex] \hline\hline
    1 &  Fiducial HOD model: Halo mass only with dark matter distribution and kinematics for satellite galaxies & MD, OR \\
    2 & Satellite galaxies have 50\% higher concentration than dark matter & MD\\
    3 & Satellite galaxies have 50\% lower concentration than dark matter & MD\\
    4 & Satellite galaxies have 50\% higher velocity dispersion than dark matter  & MD, OR \\
    5 & Satellite galaxies have 50\% lower velocity dispersion than dark matter  & MD, OR \\
    6 & The central galaxies are off-centred with a Gaussian distribution of width $0.1r_{200}$  & MD, OR \\
    7 & Assembly Bias: Central galaxies occupation is correlated with halo concentration ($A_{\rm cen}=0.3$) & MD \\
    8 & Assembly Bias: Satellite galaxies occupation is correlated with halo concentration ($A_{\rm sat}=0.3$) & MD \\
    9 & Assembly Bias: Central and Satellite galaxies occupation is correlated with halo concentration ($A_{\rm cen}=A_{\rm sat}=0.3$) & MD \\
    10 & Peculiar velocities of galaxies are scaled higher by 20\%. This should increase the growth rate by 20\% compared to the fiducial value. & MD, OR \\
    11 & Peculiar velocities of galaxies are scaled lower by 20\%. This should decrease the growth rate by 20\% compared to the fiducial value. & MD, OR \\
     \bottomrule 
     \hline\hline
    \end{tabular}
    \label{tab:HODmodels}
\end{table*}

Figure~\ref{fig:MDPL2-nonblind-pk} shows the power spectrum multipoles for the MultiDark non-blind mocks. The bottom left (right) panels show the monopole (quadruple) moment ratios with respect to the mean model. We notice that in the monopole the power spectrum ratio at large scales (small $k$) is close to 1 within 2\% except for models in which we scale the peculiar velocity (i.e. models number 10 \& 11), these models show close to 10\% difference due to the change in the Kaiser boost factor. At small scales (large $k$), we see that all SHOD models (dashed lines) have higher power in both the monopole and the quadrupole compared to HMQ models (solid lines). The models with low or high concentration for satellite galaxies show very little difference with each other (red coloured lines) within the range of scales studied and hence will not be causing any problem to RSD models. The green lines present models with low and high satellite velocity dispersion, these seem to affect the power spectrum significantly at these $k>0.2\hMpc$. The models with assembly bias show differences in the power spectrum multipoles at $k>0.2 \hMpc$ and might interfere with the growth rate measurement if the fitting scales are pushed to such small scales. Finally, magenta lines showing the most offset present models with scaled growth rate by 20\% and hence have different true cosmology and will provide a strong test of our RSD models. 
We note that the impact of baryonic physics on the matter power spectrum is considered to be important above $k \approx 0.3$ \citep{2020MNRAS.491.2424V}. But, the effect of various galaxy physics is shown at the level of 10\% by $k \approx 0.2$ in the quadrupole of galaxy power spectrum (see Figure~\ref{fig:MDPL2-nonblind-pk}). Hence the RSD analysis in this paper probes the regime affected by such beyond dark matter only physics.

\subsection{{\sc Outer Rim} Mocks}
\label{sec:OR-nonblindmocks}
Using the halo catalogue from the {\sc Outer Rim} simulation snapshot at redshift 0.86 we obtain best fit HOD parameters for SHOD, HMQ  and SFHOD model by matching the number density and large scale galaxy bias. We do not perform a detailed model fit in this case, because we are mostly interested in producing variety.

For the SHOD and HMQ models we match the observed large scale linear bias by perturbing the best fit HOD parameters obtained from MultiDark mocks. The final HOD parameters used to produce the {\sc Outer Rim} mocks are given in Table~\ref{tab:HODparameters}. We produce {\sc Outer Rim} catalogues for only 6  out of the 11 models as detailed in Table~\ref{tab:HODmodels}. The {\sc Outer Rim} halo catalogues do not come with a concentration parameter and therefore we do not include models which require this parameter. We do not use the concentration-mass relation as the true concentration has information about assembly of haloes which can not be added to a concentration simply estimated from mass. Alternatively one can fit the concentration to individual halos in {\sc Outer Rim} but due to the size and resolution of simulation this will require significant computing power which we consider out of the scope for this analysis. Also we have access to only 1 percent of particles hence such NFW fit is practically not possible.

The bottom panel of Figure~\ref{fig:wp-mock-data} shows the $w_p$ for the {\sc Outer Rim} mock catalogues. The black empty points show the measurement from the eBOSS sample with jackknife errors. The various coloured lines represent the projected correlation function measured from the mock catalogues. Solid lines are for HMQ models and dashed lines for SHOD models. The details of the different models are given in Table~\ref{tab:HODmodels}. Note that the $w_p$ at small scales is slightly underestimated. This is probably because we did not try to fit these scales and can easily be modified by allowing additional degrees of freedom to the satellite galaxies.

In \citet{avila20} there is a full account of all the mock catalogues produced with the SFHOD model and further variations. In this paper we only show the full analysis done on a subsample of SFHOD models, which complements and enhances the parameter space covered by the SHOD and HMQ models. We refer the reader to \citet{avila20}, in particular the Appendix B there, for further details. These mocks were produced by fitting the measurements of the projected correlation function and multipoles of the 3D correlation function corrected for the fibre collisions which impact the small scales of the eBOSS ELG sample. The correction was obtained using Pair-Wise Inverse probability weight \citep[PIP;][]{2017MNRAS.472.1106B,Faizan2020} method. We refer to \citet{avila20} for details of how the parameters of the models were obtained. Table~\ref{tab:SFHODmodel} describes the details of, a subset of SFHOD models has been fitted to reproduce the observed statistics, used in this work.

\begin{table}
    \centering
    \caption{ List of SFHOD models with their detailed description for the {\sc Outer Rim} simulation. The basic HOD parameters used for these models are set to $\mu=11.515, A_c=0.054, \gamma=-1.4, \sigma=0.12, A_s=0.053, \alpha=0.9, \kappa=1.0, M_c=10^{\mu-0.05}, M_1=10^{\mu+0.35}$ . Any additional degree of freedom is described below \citep[for further details see][]{avila20}.}
    \begin{tabular}{p{1.2cm} p{6cm}} \hline\hline
    {\bf Model} &  {\bf Description} \\ [0.5ex] \hline\hline
    1 &  Satellites follow a Poisson distributed, $\beta=0$ \\
    2 & Satellites follow a negative binomial distribution with $\beta=0.1$\\
    3 & Satellites follow a negative binomial distribution with $\beta=0.2$\\
    4 & Satellites follow the Next Integer from Poisson distribution, $\beta<0$\\
    5 & Satellites have an infalling velocity following a normal distribution with mean $500 \, km/s$ and standard deviation of $200 \, km/s$. \\
     \bottomrule 
     \hline\hline
    \end{tabular}
    \label{tab:SFHODmodel}
\end{table}

\subsection{RSD Results}
In this section we show the results of fitting the MultiDark and the {\sc Outer Rim} mocks with the two 
RSD models introduced in \S~\ref{sec:RSDmodels}, the TNS model (Fourier space) and the CLPT-GSRSD model (configuration space).
The analysis in Fourier and configuration space are performed as described in \S~\ref{sec:Measurements}. We note that in this paper we assume the fiducial cosmology of the mocks to be completely known and, hence, we ignore the impact that differences in fiducial cosmologies have on the results. The impact of small deviations from the fiducial cosmology on the RSD model has been discussed in our companion papers extensively \citep{de-mattia20,bautista20,gil-marin20} and shown to be small compared to the precision of our tests. We discuss the results of fitting the non-blind mocks in the following subsections.

\begin{table*}
	\centering
	\caption{Results of the redshift space distortions analysis performed on MultiDark mocks. The first set of results are for mock catalogues from the HMQ model and the second set of columns are for SHOD models. The numbers shows the best fit values and errors in the 2 least significant digits are shown in brackets. The expected values for $\alpha_\parallel$ and $\alpha_\bot$ are 1 for all models. The expected values for $f\sigma_8$ is 0.46 for models number 1-9, 0.55 for model number 10 and 0.37 for model number 11. We show results for both the Fourier space analysis with the TNS model and the configuration space analysis with the CLPT model.}
	\setlength{\tabcolsep}{3.2pt}
	\setlength{\cmidrulekern}{0.45em}
	\begin{tabular}{l ccc | ccc || ccc | ccc} 
		\hline
		 & \multicolumn{6}{c}{HMQ} & \multicolumn{6}{c}{SHOD} \\
		\cmidrule(lr){2-7}\cmidrule(lr){8-13}
		& \multicolumn{3}{c}{$P_{\ell}^{\rm TNS}$} & \multicolumn{3}{c}{$\xi_{\ell}^{\rm CLPT}$} &
		\multicolumn{3}{c}{$P_{\ell}^{\rm TNS}$} & \multicolumn{3}{c}{$\xi_{\ell}^{\rm CLPT}$} \\
		\cmidrule(lr){2-4}\cmidrule(lr){5-7}\cmidrule(lr){8-10}\cmidrule(lr){11-13}
		Model & $f\sigma_8$ & $\alpha_{\parallel}$ & $\alpha_{\bot}$  
		 & $f\sigma_8$ & $\alpha_{\parallel}$ & $\alpha_{\bot}$
		 & $f\sigma_8$ & $\alpha_{\parallel}$ & $\alpha_{\bot}$
		 & $f\sigma_8$ & $\alpha_{\parallel}$ & $\alpha_{\bot}$ \\
		\hline
1 & $0.462(29)$& $1.023(33)$& $0.988(18)$ & $0.434(48)$& $1.018(33)$& $0.975(28)$ & $0.451(28)$& $1.009(32)$& $0.975(18)$ & $0.451(40)$& $0.996(25)$& $0.976(24)$ \\ 
2 & $0.455(30)$& $1.024(35)$& $0.978(18)$ & $0.444(48)$& $1.003(31)$& $0.960(28)$ & $0.473(28)$& $0.975(30)$& $0.979(19)$ & $0.440(44)$& $1.000(28)$& $0.965(29)$ \\ 
3 & $0.465(29)$& $1.022(31)$& $0.992(18)$ & $0.426(46)$& $1.026(27)$& $0.961(26)$ & $0.470(27)$& $0.993(33)$& $0.978(17)$ & $0.469(42)$& $0.995(28)$& $0.956(26)$ \\ 
4 & $0.450(29)$& $1.008(32)$& $0.977(19)$ & $0.447(48)$& $0.994(34)$& $0.975(31)$ & $0.443(27)$& $1.011(30)$& $0.966(17)$ & $0.450(41)$& $0.988(25)$& $0.975(24)$ \\ 
5 & $0.477(30)$& $0.999(35)$& $0.976(19)$ & $0.470(50)$& $0.989(35)$& $0.970(36)$ & $0.485(27)$& $1.015(34)$& $0.996(17)$ & $0.473(46)$& $0.999(28)$& $0.976(28)$ \\ 
6 & $0.459(29)$& $0.990(33)$& $0.970(17)$ & $0.429(47)$& $1.003(29)$& $0.956(25)$ & $0.476(29)$& $0.986(32)$& $0.966(20)$ & $0.453(43)$& $0.982(27)$& $0.964(26)$ \\ 
7 & $0.451(29)$& $1.022(34)$& $0.979(17)$ & $0.458(48)$& $0.996(32)$& $0.979(27)$ & $0.466(28)$& $1.013(31)$& $0.991(18)$ & $0.444(45)$& $1.014(31)$& $0.978(28)$ \\ 
8 & $0.462(28)$& $1.010(32)$& $0.978(18)$ & $0.444(43)$& $0.998(26)$& $0.979(24)$ & $0.461(28)$& $0.994(29)$& $0.967(17)$ & $0.445(41)$& $0.989(24)$& $0.966(21)$ \\ 
9 & $0.458(28)$& $1.005(32)$& $0.970(16)$ & $0.429(45)$& $1.009(28)$& $0.965(24)$ & $0.469(28)$& $1.019(31)$& $0.992(16)$ & $0.448(39)$& $0.999(24)$& $0.976(23)$ \\ 
10 & $0.553(29)$& $1.010(29)$& $0.984(16)$ & $0.527(47)$& $1.005(28)$& $0.966(25)$ & $0.564(32)$& $1.019(31)$& $0.989(18)$ & $0.547(43)$& $1.004(29)$& $0.973(26)$ \\ 
11 & $0.348(30)$& $1.020(42)$& $0.979(18)$ & $0.373(47)$& $0.992(29)$& $0.984(23)$ & $0.375(30)$& $0.998(38)$& $0.983(20)$ & $0.353(43)$& $1.001(27)$& $0.975(26)$ \\ 
		\hline
	\end{tabular}
	\label{tab:MDPL2-RSDresults}
\end{table*}

\subsubsection{MultiDark Mocks}
\begin{figure*}
    \centering
    \includegraphics[width=1.0\textwidth]{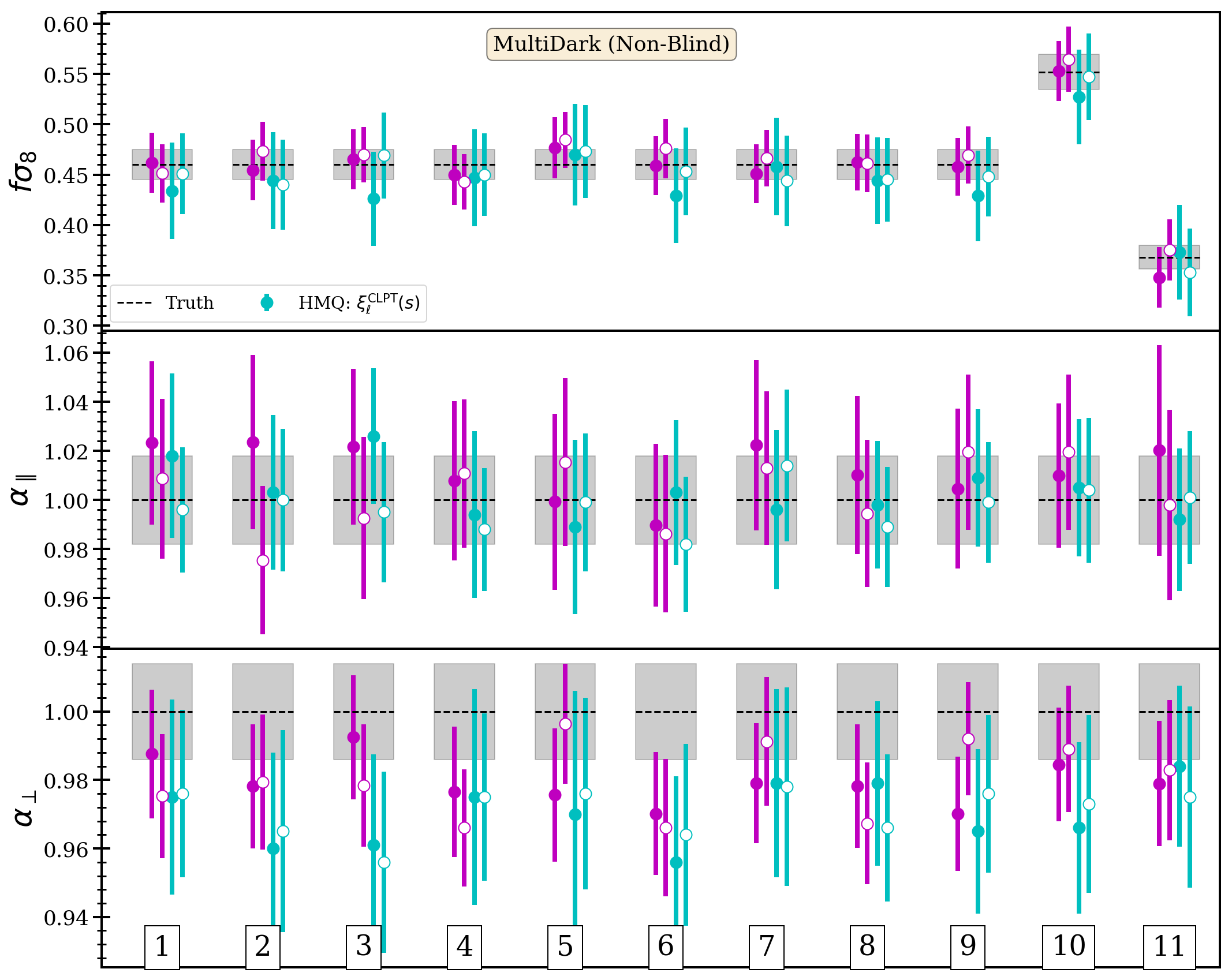}
    \caption{Results of the RSD fits to the MultiDark mocks. The three panels represents the parameters $f\sigma_8$, $\alpha_{\parallel}$ and $\alpha_{\bot}$ in the top, middle and bottom panels respectively. The x-axis shows the model number as detailed in Table~\ref{tab:HODmodels}. The magenta and cyan points correspond to the TNS and CLPT models respectively. The filled and empty points correspond to the HMQ and SHOD models for ELG. The error bars correspond to the $1\sigma$ measurement uncertainty. The black dashed lines show the true value of these parameters with 1\% bands shown in grey.}
    \label{fig:MDPL2-RSDfit}
\end{figure*}
The results of RSD fits to MultiDark mocks is given in Table~\ref{tab:MDPL2-RSDresults}.
Figure~\ref{fig:MDPL2-RSDfit} shows the results of RSD fits to the MultiDark mocks for the two RSD models considered in this paper. The top, middle and bottom panels show $f\sigma_8$, $\alpha_\parallel$ and $\alpha_\bot$ respectively. The x-axis shows the model number as detailed in Table~\ref{tab:HODmodels}. The magenta points correspond to the TNS model in Fourier space and cyan points corresponds to the CLPT-GSRSD model in configuration space. The filled points are fit to HMQ models for ELG whereas empty circles denote the fit to SHOD models for ELG. The error bars correspond to the $1\sigma$ measurement. The black dashed line shows the true value of these parameters with 1\% bands being shown in grey. 
We find that both the RSD models (TNS and CLPT-GSRSD) with the fiducial choices are in good agreement with the truth. Models numbers 2 and 3 which have significantly different small-scale clustering of satellite and model numbers 4 and 5, which have significantly different velocity dispersion of satellite galaxies do not affect the parameters obtained using the TNS and CLPT models.
Another interesting question one could ask is that, What is the impact on clustering if the central galaxies are situated away from the center of the dark matter haloes. 
The measurements from model 6, which has central galaxies away from the centre of the halo, do not show any significant bias. We consider three different kind of assembly bias in model numbers 7, 8 and 9 and find that the RSD models using large scales are again insensitive to the presence of such assembly bias in the galaxy catalogue. We also note that models numbers 10 and 11, which have a modified growth rate, can also be recovered by both RSD models without any significant bias. This has interesting confirmation that if the Universe is the same as $\Lambda$CDM model except that the growth rate is $20\%$ higher(lower) then a survey with $10\%$ statistical precision will be able to detect such effect with the models used here. In most models with MultiDark simulation, the parameter $\alpha_\bot$ seems to be underestimated by $1-1.5\sigma$. We do not detect any systematic bias in the growth rate and Alcock Paczynski scaling parameters at the level of the statistical errors of these mocks. The measurement uncertainity for these mock is a factor of two smaller than expected eBOSS ELG sample. But the detection of bias at high precision is not possible due to small volume (1 $({\rm Gpc/h})^3$) of these mocks. A more precise test for any bias in RSD models is performed using the {\sc Outer Rim} mocks.

\subsubsection{{\sc Outer Rim} Mocks}
\begin{figure*}
    \centering
    \includegraphics[width=0.45\textwidth]{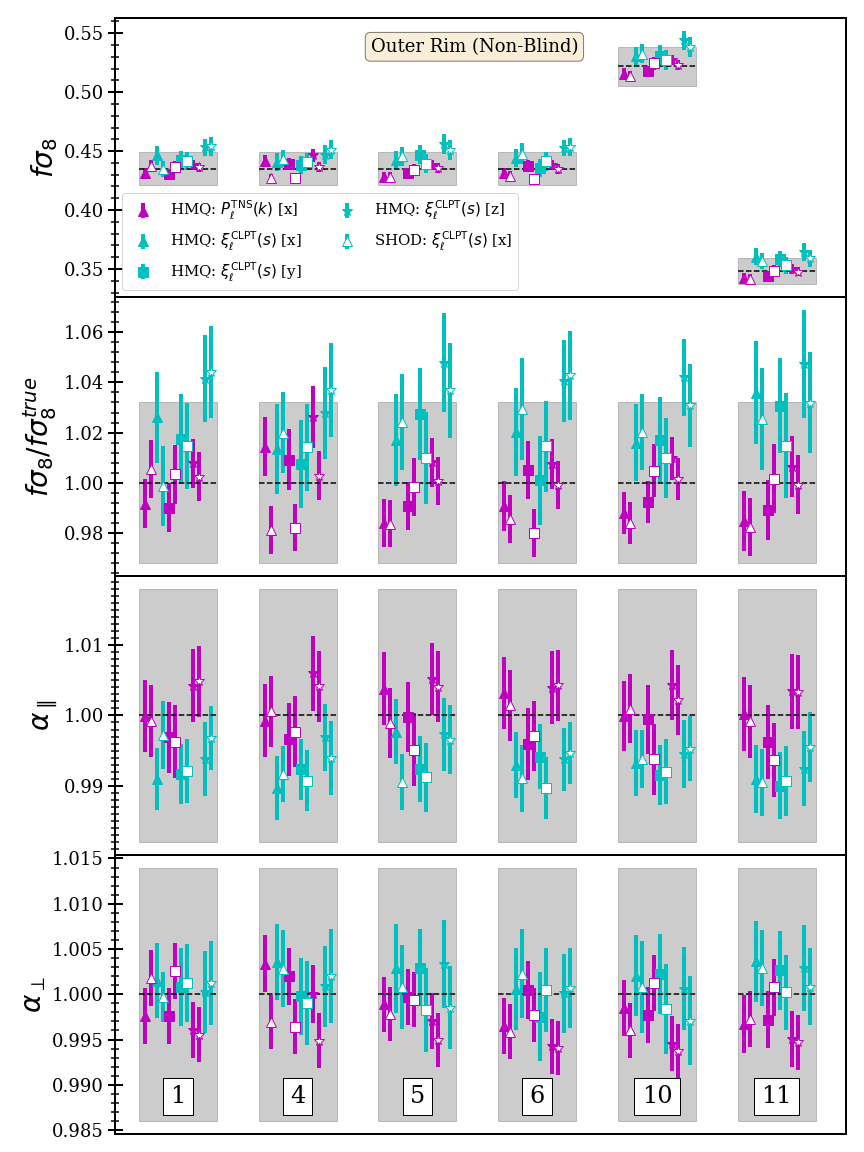}
    \includegraphics[width=0.45\textwidth]{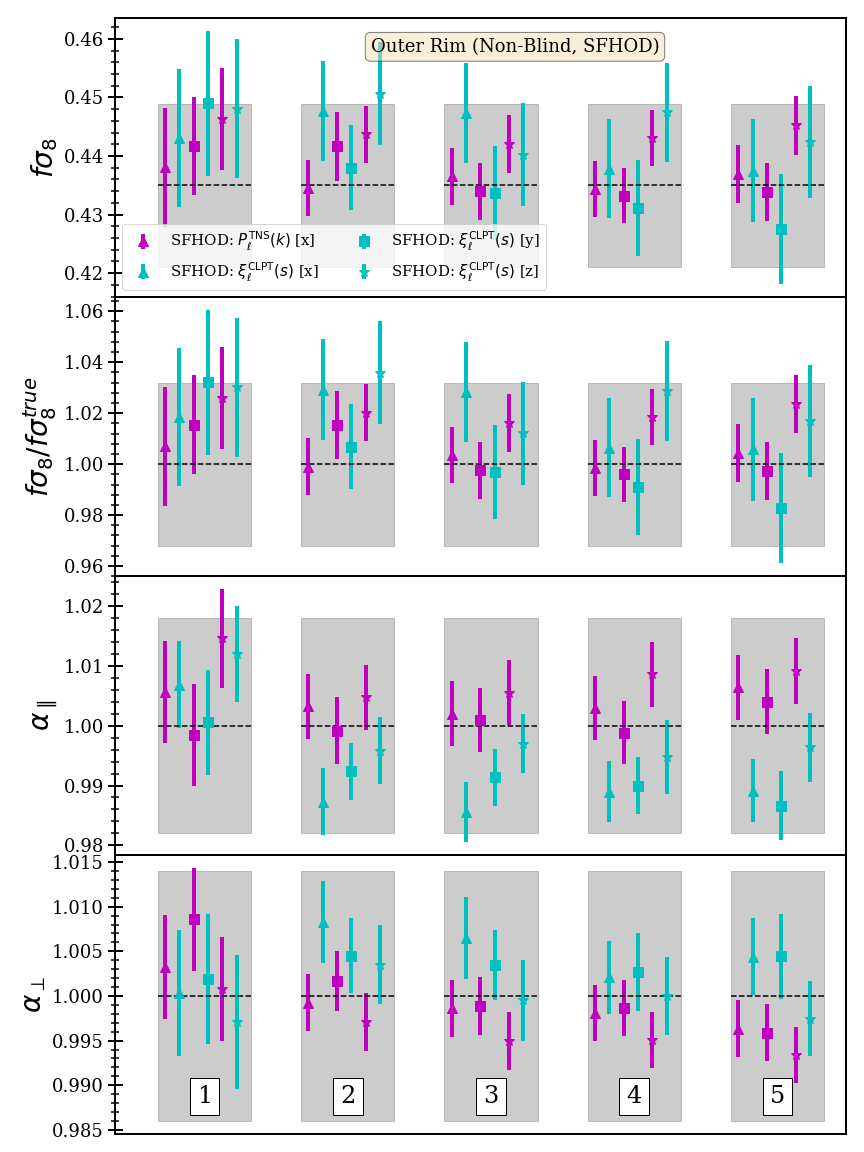}
    \caption{Results of the RSD fits to the {\sc Outer Rim} non-blind mocks. The left figure presents the SHOD and HMQ models while the right figure presents SFHOD models. The four panels in each figure represents, from top to bottom, the parameters $f\sigma_8$, $f\sigma_8/f\sigma_8^{\rm true}$, $\alpha_{\parallel}$ and $\alpha_{\bot}$. The x-axis shows the model number as detailed in Tables~\ref{tab:HODmodels} and ~\ref{tab:SFHODmodel}. The magenta and cyan points corresponds to the TNS and CLPT models respectively. In the left figure, filled and empty points corresponds to the HMQ and SHOD models for ELGs. The error bars corresponds to the $1\sigma$ measurement. The black dashed line shows the true value of these parameters with grey shaded region showing the systematic error proposed in this work. 
    }
    \label{fig:OR-RSDfit}
\end{figure*}

\begin{table*}
	\centering
	\caption{Result of redshift space distortions analysis on {\sc Outer Rim} mocks from the SHOD and HMQ models. The first set of results are for mock catalogue with HMQ model and second set of columns are for SHOD models. The numbers show the best fit values and errors in the least significant digit are shown in bracket. The expected value for $\alpha_\parallel$ and $\alpha_\bot$ is 1 for all models. The expected value for $f\sigma_8$ is 0.435 for models number 1-9, 0.522 for model number 10 and 0.348 for model number 11. We show results for both the Fourier space analysis done with the TNS model and the configuration space analysis done with the CLPT model. The x,y,z in the model name corresponds to the same mock with line-of-sight for redshift space distortions along x, y and z axis respectively.}
	\setlength{\tabcolsep}{5pt}
	\setlength{\cmidrulekern}{0.45em}
	\begin{tabular}{l ccc | ccc || ccc | ccc} 
		\hline
		 & \multicolumn{6}{c}{HMQ} & \multicolumn{6}{c}{SHOD} \\
		\cmidrule(lr){2-7}\cmidrule(lr){8-13}
		& \multicolumn{3}{c}{$P_{\ell}^{\rm TNS}$} & \multicolumn{3}{c}{$\xi_{\ell}^{\rm CLPT}$} &
		\multicolumn{3}{c}{$P_{\ell}^{\rm TNS}$} & \multicolumn{3}{c}{$\xi_{\ell}^{\rm CLPT}$} \\
		\cmidrule(lr){2-4}\cmidrule(lr){5-7}\cmidrule(lr){8-10}\cmidrule(lr){11-13}
		Model & $f\sigma_8$ & $\alpha_{\parallel}$ & $\alpha_{\bot}$  
		 & $f\sigma_8$ & $\alpha_{\parallel}$ & $\alpha_{\bot}$
		 & $f\sigma_8$ & $\alpha_{\parallel}$ & $\alpha_{\bot}$
		 & $f\sigma_8$ & $\alpha_{\parallel}$ & $\alpha_{\bot}$ \\
		\hline
1x & $0.431(4)$& $1.000(5)$& $0.998(3)$ & $0.446(7)$& $0.991(4)$& $1.002(4)$ & $0.437(5)$& $0.999(5)$& $1.002(3)$ & $0.435(6)$& $0.997(4)$& $1.000(2)$ \\ 
1y & $0.431(4)$& $0.997(5)$& $0.998(3)$ & $0.443(7)$& $0.992(4)$& $1.001(4)$ & $0.436(5)$& $0.996(5)$& $1.003(3)$ & $0.441(7)$& $0.992(4)$& $1.001(4)$ \\ 
1z & $0.438(4)$& $1.004(5)$& $0.996(3)$ & $0.453(7)$& $0.994(5)$& $1.000(4)$ & $0.436(4)$& $1.005(5)$& $0.996(3)$ & $0.454(7)$& $0.997(4)$& $1.001(4)$ \\ 
4x & $0.441(5)$& $0.999(5)$& $1.003(3)$ & $0.441(7)$& $0.990(4)$& $1.004(4)$ & $0.427(4)$& $1.001(4)$& $0.997(3)$ & $0.444(7)$& $0.992(4)$& $1.003(4)$ \\ 
4y & $0.439(5)$& $0.997(5)$& $1.002(3)$ & $0.438(7)$& $0.992(4)$& $1.000(4)$ & $0.427(4)$& $0.998(5)$& $0.996(3)$ & $0.441(7)$& $0.991(4)$& $0.999(4)$ \\ 
4z & $0.446(5)$& $1.006(5)$& $1.000(3)$ & $0.447(7)$& $0.997(4)$& $1.001(4)$ & $0.436(4)$& $1.004(5)$& $0.995(3)$ & $0.451(8)$& $0.994(5)$& $1.002(5)$ \\ 
5x & $0.428(4)$& $1.004(5)$& $0.999(3)$ & $0.442(7)$& $0.998(4)$& $1.003(4)$ & $0.428(4)$& $0.999(4)$& $0.998(2)$ & $0.446(8)$& $0.991(3)$& $1.001(4)$ \\ 
5y & $0.431(4)$& $1.000(4)$& $1.000(3)$ & $0.447(7)$& $0.992(4)$& $1.003(4)$ & $0.434(5)$& $0.995(5)$& $0.999(3)$ & $0.439(7)$& $0.991(4)$& $0.998(4)$ \\ 
5z & $0.439(4)$& $1.005(5)$& $0.997(3)$ & $0.456(8)$& $0.997(5)$& $1.003(4)$ & $0.435(4)$& $1.004(5)$& $0.995(3)$ & $0.451(8)$& $0.997(4)$& $0.999(4)$ \\ 
6x & $0.431(4)$& $1.003(5)$& $0.996(3)$ & $0.444(7)$& $0.993(4)$& $1.001(4)$ & $0.429(4)$& $1.001(5)$& $0.996(3)$ & $0.448(8)$& $0.991(4)$& $1.002(4)$ \\ 
6y & $0.437(5)$& $0.996(5)$& $1.001(3)$ & $0.435(7)$& $0.994(4)$& $0.998(4)$ & $0.426(4)$& $0.997(4)$& $0.998(3)$ & $0.441(7)$& $0.990(4)$& $1.000(4)$ \\ 
6z & $0.438(4)$& $1.004(5)$& $0.994(3)$ & $0.453(7)$& $0.994(4)$& $1.000(4)$ & $0.435(4)$& $1.004(5)$& $0.994(2)$ & $0.454(7)$& $0.995(4)$& $1.001(4)$ \\ 
10x & $0.516(4)$& $1.000(4)$& $0.998(3)$ & $0.530(8)$& $0.993(4)$& $1.002(4)$ & $0.514(4)$& $1.001(4)$& $0.996(3)$ & $0.533(7)$& $0.994(4)$& $1.001(5)$ \\ 
10y & $0.518(4)$& $0.999(4)$& $0.998(3)$ & $0.531(8)$& $0.992(4)$& $1.002(4)$ & $0.525(5)$& $0.994(5)$& $1.001(3)$ & $0.527(8)$& $0.992(4)$& $0.998(4)$ \\ 
10z & $0.527(4)$& $1.004(5)$& $0.995(3)$ & $0.544(7)$& $0.995(4)$& $1.001(4)$ & $0.523(4)$& $1.002(4)$& $0.994(2)$ & $0.538(8)$& $0.995(4)$& $0.997(4)$ \\ 
11x & $0.343(4)$& $1.000(5)$& $0.997(3)$ & $0.361(7)$& $0.991(4)$& $1.004(4)$ & $0.342(4)$& $0.999(5)$& $0.997(3)$ & $0.357(7)$& $0.990(4)$& $1.003(4)$ \\ 
11y & $0.344(4)$& $0.996(5)$& $0.997(3)$ & $0.359(6)$& $0.990(4)$& $1.003(4)$ & $0.349(4)$& $0.994(5)$& $1.001(3)$ & $0.353(7)$& $0.991(5)$& $1.000(4)$ \\ 
11z & $0.350(4)$& $1.003(5)$& $0.995(3)$ & $0.365(7)$& $0.992(5)$& $1.003(4)$ & $0.348(4)$& $1.003(5)$& $0.995(3)$ & $0.359(7)$& $0.996(5)$& $1.001(4)$ \\
		\hline
	\end{tabular}
	\label{tab:OR-RSDresults}
\end{table*}

\begin{table}
	\centering
	\caption{Result of redshift space distortions analysis on SFHOD mocks using the {\sc Outer Rim} simulation. We provide the best fit values and errors in the least significant digits are shown in bracket. Results are given for both Fourier space analysis with TNS model and configuration space analysis with CLPT model. The x,y,z in the model name corresponds to the same mock with line-of-sight for redshift space distortions along x, y and z axis respectively. 
	}
	\setlength{\tabcolsep}{3pt}
	\setlength{\cmidrulekern}{0.45em}
	\begin{tabular}{l ccc | ccc} 
		& \multicolumn{3}{c}{$P_{\ell}^{\rm TNS}$} & \multicolumn{3}{c}{$\xi_{\ell}^{\rm CLPT}$} \\
		\cmidrule(lr){2-4}\cmidrule(lr){5-7}
		Model & $f\sigma_8$ & $\alpha_{\parallel}$ & $\alpha_{\bot}$  
		 & $f\sigma_8$ & $\alpha_{\parallel}$ & $\alpha_{\bot}$ \\
		\hline
1x & $0.438(10)$& $1.006(8)$& $1.003(5)$ & $0.443(11)$& $1.007(7)$& $1.000(7)$ \\ 
1y & $0.442(8)$& $0.998(8)$& $1.009(5)$ & $0.449(12)$& $1.001(8)$& $1.002(7)$ \\ 
1z & $0.446(8)$& $1.015(8)$& $1.001(5)$ & $0.448(11)$& $1.012(8)$& $0.997(7)$ \\ 
2x & $0.435(4)$& $1.003(5)$& $0.999(3)$ & $0.448(8)$& $0.987(5)$& $1.008(4)$ \\ 
2y & $0.442(5)$& $0.999(5)$& $1.002(3)$ & $0.438(7)$& $0.992(4)$& $1.005(4)$ \\ 
2z & $0.444(4)$& $1.005(5)$& $0.997(3)$ & $0.451(8)$& $0.996(5)$& $1.004(4)$ \\ 
3x & $0.437(4)$& $1.002(5)$& $0.999(3)$ & $0.447(8)$& $0.986(4)$& $1.007(4)$ \\ 
3y & $0.434(4)$& $1.001(5)$& $0.999(3)$ & $0.434(7)$& $0.991(4)$& $1.003(3)$ \\ 
3z & $0.442(4)$& $1.005(5)$& $0.995(3)$ & $0.440(8)$& $0.997(4)$& $1.000(4)$ \\ 
4x & $0.434(4)$& $1.003(5)$& $0.998(3)$ & $0.438(8)$& $0.989(5)$& $1.002(4)$ \\ 
4y & $0.433(4)$& $0.999(5)$& $0.999(3)$ & $0.431(8)$& $0.990(4)$& $1.003(4)$ \\ 
4z & $0.443(4)$& $1.009(5)$& $0.995(3)$ & $0.447(8)$& $0.995(6)$& $1.000(4)$ \\ 
5x & $0.437(4)$& $1.006(5)$& $0.996(3)$ & $0.438(8)$& $0.989(5)$& $1.004(4)$ \\ 
5y & $0.434(4)$& $1.004(5)$& $0.996(3)$ & $0.428(9)$& $0.987(5)$& $1.004(4)$ \\ 
5z & $0.445(5)$& $1.009(5)$& $0.993(3)$ & $0.442(9)$& $0.996(5)$& $0.997(4)$ \\ 
			\hline
	\end{tabular}
	\label{tab:OR-SFHOD-RSDresults}
\end{table}

Figure~\ref{fig:OR-RSDfit} shows the results of the RSD fits to the {\sc Outer Rim} mocks for the two RSD models considered in this paper. The left figure presents the SHOD and HMQ models and the right figure SFHOD models. 
The four panels in each figure presents, from top to bottom, the parameters $f\sigma_8$, $f\sigma_8/f\sigma_8^{\rm true}$, $\alpha_{\parallel}$ and $\alpha_{\bot}$. 
The x-axis in this figure shows the model number as detailed in Table~\ref{tab:HODmodels} and ~\ref{tab:SFHODmodel}. The magenta points correspond to the TNS model in Fourier space and the cyan points to the CLPT-GSRSD model in configuration space. In the left figure, filled symbols show the fit of HMQ models to ELGs and empty points show the fit of SHOD models. The error bars corresponds to the $1\sigma$ measurement. The black dashed line shows the true value of these parameters with the grey shaded region showing the systematic error proposed in this work. The large volume of these mocks  (27 $({\rm Gpc/h})^3$) results in very small statistical uncertainties. The statistical errors in these mocks is less than $2\%$ for $f\sigma_8$ and less than $1\%$ for $\alpha_{\parallel}$ and $\alpha_{\bot}$.  We find that both RSD models (TNS and CLPT-GSRSD) with the fiducial cosmological choices are in good agreement with the truth at the level of the statistical precision 
of these mocks. The uncertainity in these mocks is about 1/10$^th$ of the eBOSS ELG sample, hence this should provide a reliable estimate of theoretical systematic errors for the purpose of the eBOSS ELGs sample.

It is interesting to ask whether baryonic effects can bias such cosmological measurements when performed at percent level. There are several different ways in which baryonic physics can impact the galaxy samples.
Several aspects of the complex baryonic processes can lead to incomplete sample of ELG galaxies compared to a mass selected sample. This can be related to galaxy quenching, expulsion of cold gas from hot haloes, outflows from AGNs, supernovae events, etc. 
Therefore the lack of systematic biases in the measured parameters in SHOD vs HMQ model (this encapsulate the mass incompleteness in a different way) is a remarkable success of TNS and CLPT-GSRSD model. 
This indicates that despite the details on how the mass incompleteness is modelled affecting the small-scale clustering, the RSD models when using relatively large scales can provide unbiased measurements of the cosmological parameters at a percent level. The effects of various dynamical process can possibly increase or decrease the velocity dispersion of satellites. Model number 4 (i.e SHOD4, HMQ4) and 5 (i.e. SHOD5, HMQ5) tests for such effects and shows no significant bias in the RSD parameters. Another additional feature, the observed galaxy catalogue may have, is that the central galaxies are shifted from the centre of the dark matter haloes. Model  number 6 (i.e. HMQ6 and SHOD6) aims to mimic this effect and RSD fits are again unbiased. 

The SFHOD models also show an unbiased measurement of RSD parameters (see right panel of Figure~\ref{fig:OR-RSDfit}). The SFHOD model number 1 assumes that  satellite galaxies follow a Poisson distribution whereas model numbers 2 and 3 use a negative binomial distributions with $\beta=0.1$ and $\beta=0.2$ respectively. The SFHOD model number 4 assumes a next integer from Poisson distribution for satellites. These models have different small-scale physics and hence different Finger-of-God effects which can arise due to baryon physics (note that a next integer distribution has been reported in SAMs, which are not directly accounting for hydrodynamical interactions). The SFHOD model number 5 introduces an infalling velocity component to satellite galaxies motivated by behaviour of model galaxies in \citet{2018MNRAS.475.2530O}. All of the five SFHOD models do not show any significant systematic bias in the RSD parameters beyond the statistical uncertainty of the mocks (see figure ~\ref{fig:OR-RSDfit}).

We have shown how the results obtained vary for each mock galaxy catalogue when projecting the RSD along different los. The scatter that is seen between the different los looks consistent with the errors, and we mitigate it by averaging together the three measurements. In the mocks that are produced for the quasar sample, a much larger scatter is seen in the $f\sigma_8$ measurements for different los \citep{smith20}. This is investigated in more detail in Smith et al. (in prep), where it is shown that the scatter in $f\sigma_8$ is larger for tracers with a larger linear bias. Since measurements of the quadrupole (and hence $f\sigma_8$) are anti-correlated, large gains in the precision can be made by taking the mean of the three los, which is greater than what would be expected if the volume was increased by a factor of 3.

Overall we show in figure~\ref{fig:OR-RSDfit} along with Table~\ref{tab:OR-RSDresults} and~\ref{tab:OR-SFHOD-RSDresults} that the way the HOD models encapsulate different  baryonic physics for ELGs do not bias the RSD parameters. This is true for both TNS and CLPT-GSRSD models to few percent precision. The RSD models are unbiased when limited to using $k_{\rm max}=0.2\hMpc$ for Fourier space and $s_{\rm min}=32 \,h^{-1}{\rm Mpc} $ in configuration space. This is a remarkable success of the perturbation theory schemes against such wide variety of galaxy formation models along with various forms of halo mass incompleteness. This result is encouraging while looking forward to very precise measurements in the future. 

\section{Blind Mock Challenge}
\label{sec:blind}
In this section we describe the blind part of our mock challenge. The main focus of this measurement has been testing the ability of perturbation theory based redshift space distortions models to obtain unbiased growth rate ($f\sigma_8$) measurements. Therefore, we create a new set of mocks using the {\sc Outer Rim} simulation with a blind true growth rate and provide that to participants for analysis. The blind mocks are created by scaling the halo velocity linearly resulting in scaling of true growth rate. We create 6 blind mocks in this paper, three using SHOD models to populate ELGs and another three using HMQ models. These mocks are analysed using the same method and scales as for the analysis on non-blind mocks, as described in detail in \S~\ref{sec:Measurements}. The number density for blind mocks is set to the $2 \times 10^{-4} \mpcohcub$, this is close to the mean number density of the eBOSS ELG sample. We have created 30 realisations for each of the three HMQ models and 40 for each of the three SHOD models. These realisations are created from the same halo catalogue but sub-sampling randomly a distinct set of haloes for each realisation. The number of realisations is set by the total number density obtained for the full halo catalogue based on the HOD model. The error quoted in the Table~\ref{tab:OR-RSDresults-blind} for the blind mocks is the scatter in the dispersion of these 30(40) realisations for the HMQ(SHOD) models.

The six blind mocks use the same underlying halo catalogue from {\sc Outer Rim} as the non-blind mocks presented in the previous section and hence the underlying  cosmological parameters were known to everyone. An analysis with blind cosmological parameters are left for the future. In this paper we focus on the ability to constrain the growth rate, rather than the full cosmology.
We generate three models for each of HMQ and SHOD ELG models using the underlying parameters given in Table~\ref{tab:HODparameters}. We scale the growth rate by 0.5 for the blind mock number 1, by 0.75 for blind mock number 2 and by 1.0 for number 3. The RSD along each of the axis of cubic box were applied as indicated in the Table~\ref{tab:OR-RSDresults-blind} by $x,y$ and $z$ with the model number. 
These shifts were kept blinded until we finalised all the plots and tables for this paper. The shifts in the growth rate are at $30\sigma$ and $15\sigma$ level assuming $~1.6\%$ statistical uncertainty in measurement of growth rate. The shifts are set to such large values in order to study whether the model can obtain an unbiased estimate of growth rate at percent level precision despite it being far away from the default value.

\subsection{Blind Mocks results}


\begin{figure*}
    \centering
    \includegraphics[width=1.0\textwidth]{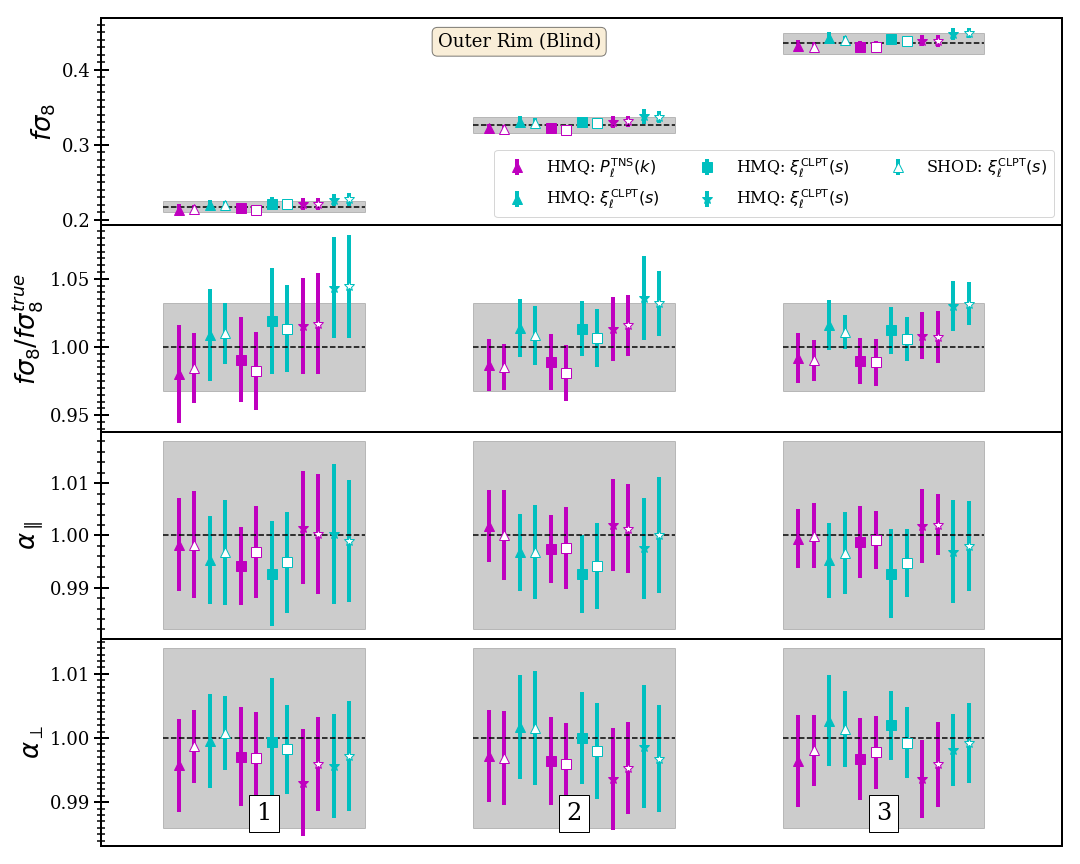}
    \caption{Results of RSD fits to the {\sc Outer Rim} blind mocks. From top to bottom, the four panels presents the parameters $f\sigma_8$,$f\sigma_8/f\sigma_8^{\rm true}$, $\alpha_{\parallel}$ and $\alpha_{\bot}$. The x-axis shows the three  different blind models with traingle, square and star marker for RSD realizations along x,y and z axis respectively (see Table~\ref{tab:OR-RSDresults-blind} for values).
    The magenta and cyan points correspond to the TNS and CLPT models respectively. The filled and empty points correspond to the HMQ and SHOD models for ELGs. The error bars correspond to the $1\sigma$ measurement uncertainty. The black dashed lines show the true value of these parameters with grey shaded region showing the systematic error proposed in this work. It is clear from the top panel that our choice of blind mocks cover a wide range of growth rate which is consistently recovered by the two RSD models.}
    \label{fig:OR-RSD-blind}
\end{figure*}

\begin{figure*}
    \centering
    \includegraphics[width=1.0\textwidth]{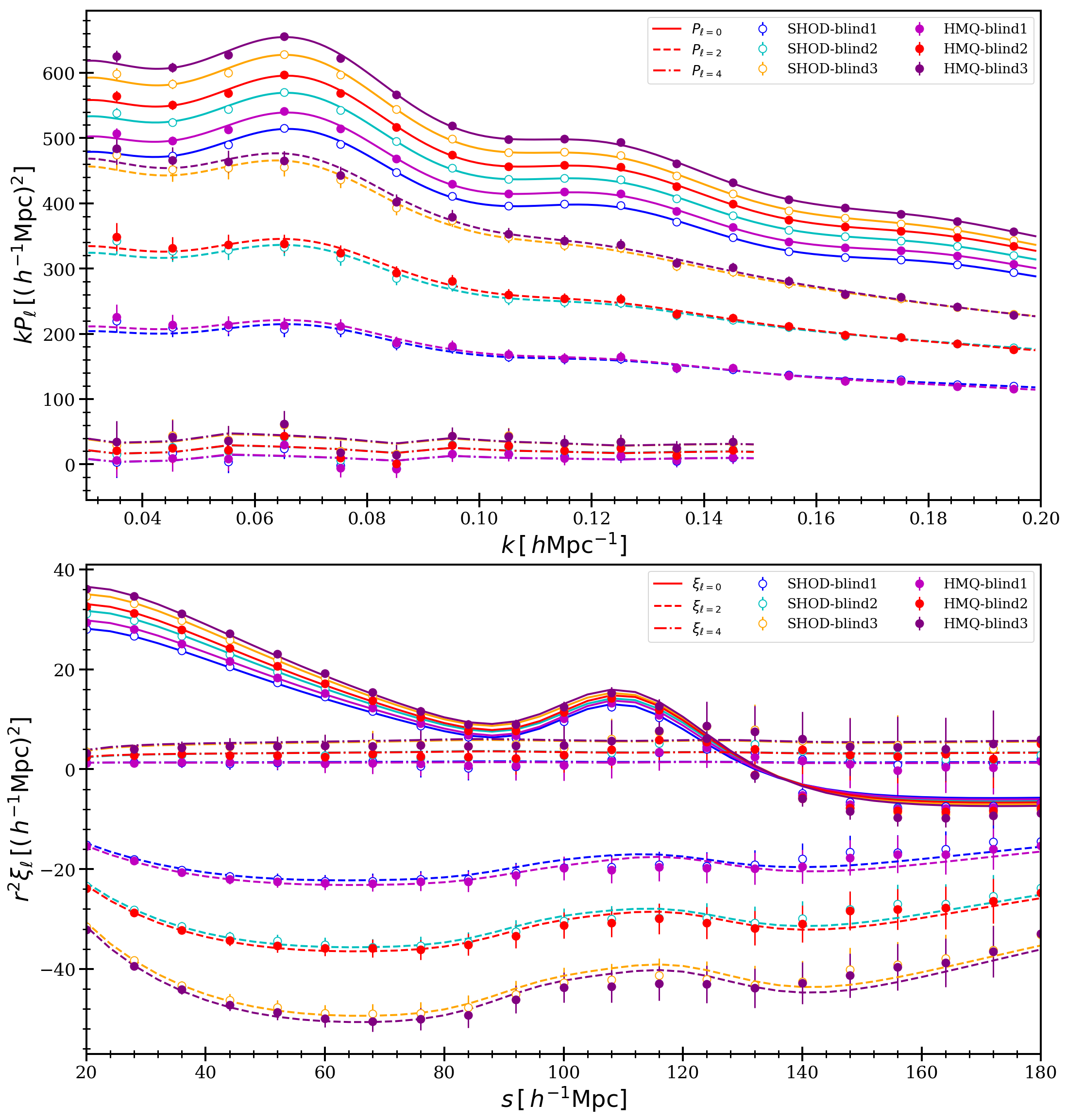}
    \caption{Clustering measurement for the blind mocks along with the best fit models. The top panel shows the power spectrum multipoles and the bottom panel the correlation function multipoles. The measurements from the mocks along with their error are shown with circles, whereas the best fit models are shown with lines. The mocks from the model with quenching (i.e. HMQ) are shown with filled symbols and those from the model without quenching (i.e. SHOD) are shown with empty symbols. The solid, dashed and dotted-dashed lines are for the Legendre moments with $\ell=0$, $\ell=2$ and $\ell=4$ respectively. The power spectrum uses TNS whereas the correlation function uses CLPT-GSRSD to model the redshift space distortions. Both models describe the mocks remarkably successfully.}
    \label{fig:pk-xi-blind-wbfit}
\end{figure*}

\begin{table*}
	\centering
	\caption{Results of the redshift space distortions analysis on the {\sc Outer Rim} blind mocks. The first set of results are for mock catalogues from the HMQ models and the second set of columns are from SHOD models. The numbers show the best fit values and errors in the least significant digits are shown within bracket.
	We show results for both the Fourier space analysis with the TNS model and the configuration space analysis with the CLPT model. The x,y,z in the model name corresponds to the same mock with line-of-sight for redshift space distortions along x, y and z axis respectively.}
	\setlength{\tabcolsep}{5pt}
	\setlength{\cmidrulekern}{0.45em}
	\begin{tabular}{l ccc | ccc || ccc | ccc} 
		\hline
		 & \multicolumn{6}{c}{HMQ} & \multicolumn{6}{c}{SHOD} \\
		\cmidrule(lr){2-7}\cmidrule(lr){8-13}
		& \multicolumn{3}{c}{$P_{\ell}^{\rm TNS}$} & \multicolumn{3}{c}{$\xi_{\ell}^{\rm CLPT}$} &
		\multicolumn{3}{c}{$P_{\ell}^{\rm TNS}$} & \multicolumn{3}{c}{$\xi_{\ell}^{\rm CLPT}$} \\
		\cmidrule(lr){2-4}\cmidrule(lr){5-7}\cmidrule(lr){8-10}\cmidrule(lr){11-13}
		Model & $f\sigma_8$ & $\alpha_{\parallel}$ & $\alpha_{\bot}$  
		 & $f\sigma_8$ & $\alpha_{\parallel}$ & $\alpha_{\bot}$
		 & $f\sigma_8$ & $\alpha_{\parallel}$ & $\alpha_{\bot}$
		 & $f\sigma_8$ & $\alpha_{\parallel}$ & $\alpha_{\bot}$ \\
		\hline
1x & $0.213(7)$& $0.998(8)$& $0.996(7)$ & $0.219(7)$& $0.995(8)$& $1.000(7)$ & $0.214(5)$& $0.998(10)$& $0.999(5)$ & $0.220(4)$& $0.997(10)$& $1.001(5)$ \\ 
1y & $0.215(6)$& $0.994(7)$& $0.997(7)$ & $0.222(8)$& $0.993(10)$& $0.999(9)$ & $0.214(6)$& $0.997(8)$& $0.997(7)$ & $0.220(6)$& $0.995(9)$& $0.998(6)$ \\ 
1z & $0.221(7)$& $1.002(10)$& $0.993(8)$ & $0.227(8)$& $1.000(13)$& $0.996(8)$ & $0.221(8)$& $1.000(11)$& $0.996(7)$ & $0.227(8)$& $0.999(11)$& $0.997(8)$ \\ 
2x & $0.322(6)$& $1.002(6)$& $0.997(7)$ & $0.331(6)$& $0.997(7)$& $1.002(8)$ & $0.321(5)$& $1.000(8)$& $0.997(7)$ & $0.329(7)$& $0.997(8)$& $1.002(8)$ \\ 
2y & $0.323(6)$& $0.997(6)$& $0.996(6)$ & $0.331(6)$& $0.993(7)$& $1.000(7)$ & $0.320(6)$& $0.998(7)$& $0.996(6)$ & $0.328(7)$& $0.994(8)$& $0.998(7)$ \\ 
2z & $0.330(7)$& $1.002(8)$& $0.994(7)$ & $0.338(10)$& $0.998(9)$& $0.999(9)$ & $0.331(7)$& $1.001(8)$& $0.995(7)$ & $0.337(7)$& $1.000(11)$& $0.997(8)$ \\ 
3x & $0.431(7)$& $0.999(5)$& $0.996(7)$ & $0.442(7)$& $0.995(7)$& $1.003(7)$ & $0.431(6)$& $1.000(6)$& $0.998(5)$ & $0.440(5)$& $0.997(7)$& $1.001(5)$ \\ 
3y & $0.431(7)$& $0.999(6)$& $0.997(6)$ & $0.440(7)$& $0.993(8)$& $1.002(5)$ & $0.430(7)$& $0.999(5)$& $0.998(5)$ & $0.438(6)$& $0.995(6)$& $0.999(5)$ \\ 
3z & $0.439(7)$& $1.002(7)$& $0.994(6)$ & $0.448(7)$& $0.997(9)$& $0.998(5)$ & $0.438(8)$& $1.002(5)$& $0.996(6)$ & $0.449(6)$& $0.998(8)$& $0.999(6)$ \\
\hline
	\end{tabular}
	\label{tab:OR-RSDresults-blind}
\end{table*}

We present the result of analysing the blind mocks from {\sc Outer Rim} simulation in Figure~\ref{fig:OR-RSD-blind} and~\ref{fig:pk-xi-blind-wbfit}. Table~\ref{tab:OR-RSDresults-blind} summarises the parameter constraints.
Figure~\ref{fig:OR-RSD-blind} shows the results of the RSD fits using the two models, TNS and CLPT. From top to bottom, the four panels represents the parameters $f\sigma_8$,$f\sigma_8/f\sigma_8^{\rm true}$, $\alpha_{\parallel}$ and $\alpha_{\bot}$. The error bars corresponds to the $1\sigma$ measurement.
It is clear from the top panel in Figure~\ref{fig:OR-RSD-blind} that our choice of blind mocks cover a wide range of growth rates. These are consistently recovered by the two RSD models with a precision close to the percent level. The large volume of these blind mocks result in very small statistical uncertainties. The mean one sigma statistical error in growth rate are $1.6\%$, $2.2\%$, and $3.2\%$ for model number 3, 2 and 1 respectively (see Table~\ref{tab:OR-RSDresults-blind}).
The mean one sigma statistical errors for $\alpha_{\parallel}$ is $0.9\%$  and for $\alpha_{\bot}$ is $0.7\%$.
We note that the uncertainties in the configuration space analysis using CLPT are fairly close to those obtained in Fourier space using the TNS model. 
We also note that the mean deviation of the parameters measured from the true values are $\left[0.6, 0.2, 0.6\right]\times \sigma_{\rm stat}$ for $f\sigma_8$, $\alpha_{\parallel}$ and $\alpha_{\bot}$ respectively for the TNS model. Whereas the configuration space analysis results in mean deviation of $\left[0.9, 0.4, 0.2\right]\times \sigma_{\rm stat}$  for $f\sigma_8$, $\alpha_{\parallel}$ and $\alpha_{\bot}$ respectively. 
We also note that the statistical errors for the blind mocks are probably slightly underestimated given they are coming from a small number of realisations sampled from the same halo catalogue. Therefore, we do not detect any systematic bias in the blind mock challenge for the TNS and CLPT models at the statistical precision of these mocks. Note that this remarkable success of the TNS and CLPT models might be partially driven by the fact that we kept a conservative cut in scale to limit the impact of non-linear growth of dark matter and baryonic physics on our measurements. 
It remains to be seen how far one can push in non-linear scales when analysing the mocks with one-tenth of the statistical error used in this work as this will be the typical requirement of future surveys. 

In order to illustrate the accuracy and success of the redshift space distortions models we show measurements from the blind mocks along with the best fit models in Figure ~\ref{fig:pk-xi-blind-wbfit}. The top panel shows the power spectrum multipoles and the bottom panel is for the correlation function ones. The measurements from the mocks along with their error are shown with circles whereas the best fit models are shown with lines. The mocks from models with quenching (i.e. HMQ) are shown with filled symbols and those from models without quenching (i.e. SHOD) are shown with empty symbols. The solid, dashed and dotted-dashed lines are for Legendre moments with $\ell=0$, $\ell=2$ and $\ell=4$ respectively. The power spectrum uses TNS whereas the correlation function uses CLPT-GSRSD to model the redshift space distortions. The SHOD blind model number 1,2 and 3 are shown with blue cyan and orange colours respectively. Whereas HMQ blind model number 1,2 and 3 are shown with magenta, red and purple colours respectively. For the model to work, the same coloured line (best fit from theory) should go through the same coloured points (measurements from the blind mocks). Taking this into account, we can conclude that both models describe the blind mocks remarkably successfully.
\section{Systematic errors}
\label{sec:sys}

The final aim of this paper is to provide the theoretical systematic uncertainties related to the modelling of the clustering of the eBOSS ELG sample. Therefore, we provide estimates of systematic error in the measurements of $f\sigma_8$, $\alpha_{\parallel}$ and $\alpha_{\bot}$ for the TNS model in Fourier space and the CLPT model in configuration space. This systematic error asses the impact of galaxy formation physics. 
In particular we consider impact of quenching mechanism, assembly bias, in falling of satellite galaxies, satellites having different concentration and velocity dispersion compared to dark matter through various mock catalogues.

We measure three quantities for each of the RSD parameters $f\sigma_8$, $\alpha_{\parallel}$ and $\alpha_{\bot}$, and type of HOD model described in~\S~\ref{sec:ELGmodels} to asses the systematic bias, as given below: 

\begin{align}
\mu_{\rm sys} &= \left< \left| x-x_{\rm true} \right| \right>  \label{eq:musys} \, ,\\
\sigma_{\rm sys}^2 &=  \left< \left(x-x_{\rm true} \right)^2\right>- \left< x-x_{\rm true} \right>^2  \label{eq:sigsys} \, ,\\
\sigma_{\rm stat} &= \mathrm{statistical\, error\, from\, fit} \, .  \label{sigstat}
\end{align}
Above, x represents one of the RSD parameters (i.e. $f\sigma_8$, $\alpha_{\parallel}$, $\alpha_{\bot}$) and averages are taken over all the mocks in a given type of HOD model. 
The parameters $\mu_{\rm sys}$ represent the mean systematic shift from the true value, $\sigma_{\rm sys}$ represents the rms of the systematic shift and $\sigma_{\rm stat}$ represents the  statistical error. 
We consider a systematic bias is significant only if $\mu_{\rm sys} > 2\sigma_{\rm stat}$.
Assuming Gaussian statistics for the systematic errors, this requirement implies that we only detect a systematic bias if statistically there is only a 5\% chance to explain the distance from the measured parameter to the truth.

\begin{figure}
    \centering
    \includegraphics[width=0.49\textwidth]{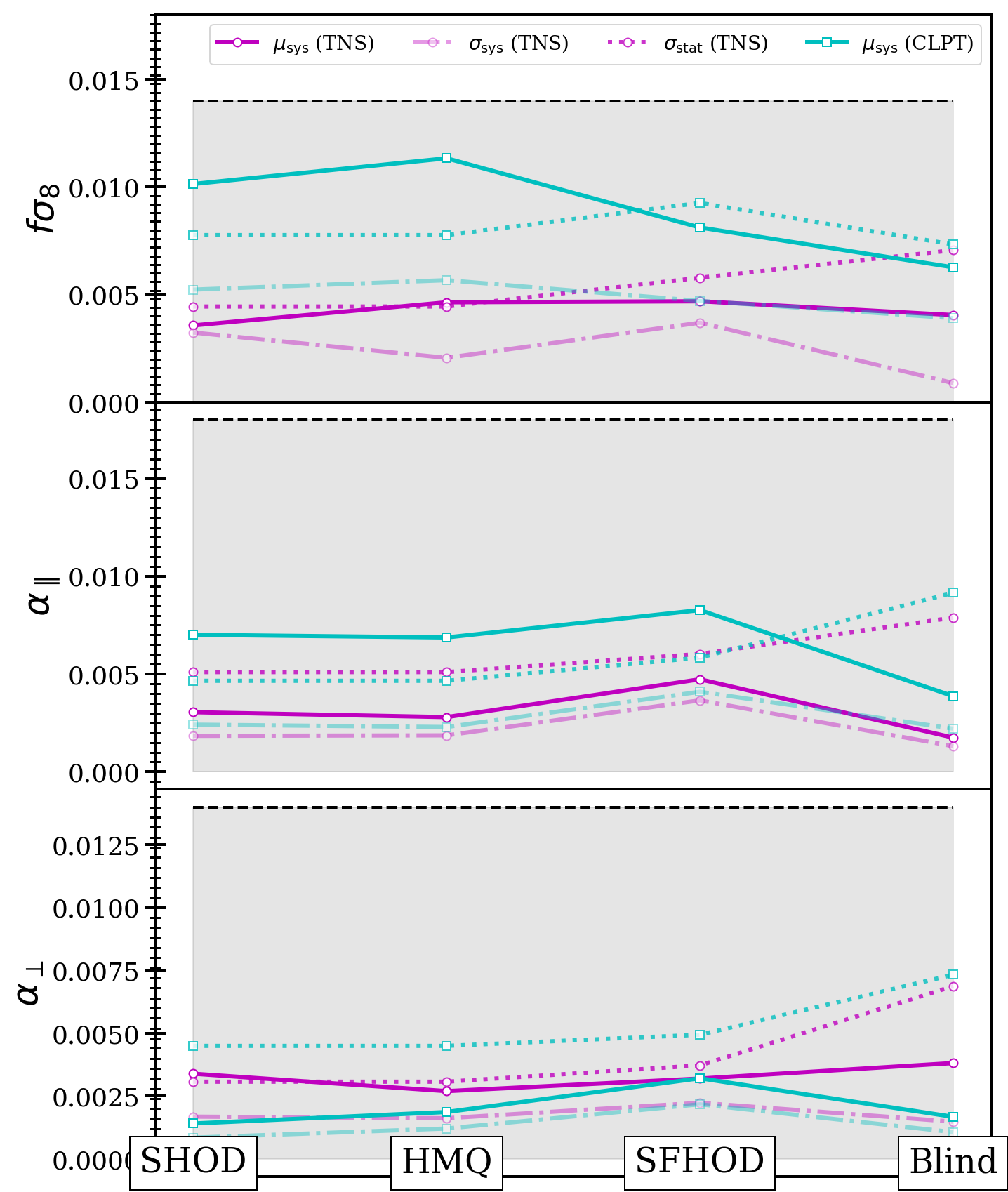}
    \caption{Comparison of systematic and statistical absolute errors for $f\sigma_8$ (top), $\alpha_{\parallel}$ (middle) and $\alpha_{\bot}$ (bottom). The x-axis shows the results for different kind of mocks. In each panel the solid line is for $\mu_{\rm sys}$ representing mean systematic shift, dotted-dashed line is for $\sigma_{\rm sys}$ representing the rms of mean systematic shift from true values of the parameter and dashed line is for $\sigma_{\rm sys}$ representing statistical errors. The magenta colour is for the TNS model and the cyan colour for CLPT. The shaded region with black dashed line shows the theoretical systematic error proposed in this work. We note that a detection of systematic bias as per our definition will mean that coloured solid lines above twice the value of the corresponding colored dashed lines. Therefore this figure illustrates that for both RSD model and all four type of mock catalogues we do not detect any systematic bias.
    }
    \label{fig:sys-err}
\end{figure}

The measurement of these three parameters ($\mu_{\rm sys}$, $\sigma_{\rm sys}$, $\sigma_{\rm stat}$) are shown in Figure~\ref{fig:sys-err} and given in Table~\ref{tab:sys-error}. Table~\ref{tab:sys-error} lists the systematic error in each of the three RSD parameters for the two RSD models, TNS and CLPT, and 4 different HOD model categories, SHOD, HMQ, SFHOD and Blind (all of the Blind models) based on the {\sc Outer Rim} simulation. We do not provide these values for mocks based on the MultiDark simulation as the statistical errors in those are much larger. The systematic shift ($\mu_{\rm sys}$) for the TNS model in $f\sigma_8$ is $0.004$, $0.005$, $0.005$ and $0.004$ for the SHOD, HMQ, SFHOD and Blind mocks respectively. These systematic shifts are either smaller or at the level of the statistical errors. Although $\mu_{\rm sys}$ for the CLPT model in $f\sigma_8$ for SHOD ($0.01$) and HMQ ($0.011$) is slightly larger than the statistical errors, this difference is not statistically significant. The systematic shift ($\mu_{\rm sys}$) for the TNS model in $\alpha_{\parallel}$ and $\alpha_{\bot}$ is always smaller than the corresponding $\sigma_{\rm stat}$ for all the four mocks, except for $\alpha_{\bot}$ in SHOD model. 
The systematic shift for the CLPT model in  $\alpha_{\bot}$ is smaller than $\sigma_{\rm stat}$ for all four mocks but the one in $\alpha_{\parallel}$ are larger than $\sigma_{\rm stat}$. But this shift in $\alpha_{\parallel}$ does n't cross our requirement of significant systematic(i.e. $\mu_{\rm sys}>2 \times \sigma_{\rm stat}$).

The numbers given in Table~\ref{tab:sys-error} for the systematic errors can be visualised in Figure ~\ref{fig:sys-err}. The top, middle and bottom panels in this plot correspond to the absolute errors for $f\sigma_8$, $\alpha_{\parallel}$ and $\alpha_{\bot}$ respectively. The x-axis shows the results for different categories of mocks as indicated at the bottom panel. In Figure ~\ref{fig:sys-err}, solid lines are for $\mu_{\rm sys}$, the mean systematic shift, dotted-dashed lines are for $\sigma_{\rm sys}$, the standard deviation of mean systematic shift, and dotted lines are for $\sigma_{\rm stat}$, the statistical errors. The magenta colour shows results from the power spectrum analysis using the TNS model and the cyan colour shows results from the correlation function analysis using the CLPT model. In this work, only solid lines above twice the value of the dotted lines do imply the existence of a significant systematic bias. Therefore, Figure~\ref{fig:sys-err} illustrates that for both RSD models and all four categories of mock catalogues we do not detect any systematic bias. 

The error in the three RSD parameters for blind mocks is fairly close between the Fourier space analysis using TNS and the configuration space analysis using CLPT. 
We remind readers that the Fourier space analysis uses $k_{\rm max}=0.2$, whereas the configuration space analysis uses $s_{\rm min}=32$. The two scale cuts are equivalent if we related the Fourier conjugates as $k_{\rm max}=2\pi/s_{\rm min}$. Hence the two analysis shows similar level of information from the two-point clustering of galaxies.

We take a conservative choice and suggest using twice the statistical error in the blind mocks as the theoretical systematic. 
We propose the following theoretical systematic errors, common to the two RSD models, TNS and CLPT: 0.0146 for $f\sigma_8$, 0.0184 for $\alpha_{\parallel}$,  and 0.0146 for $\alpha_{\bot}$. This results in a theoretical systematic error budget of $3.3\%$,  $1.8\%$ and $1.5\%$ in $f\sigma_8$, $\alpha_{\parallel}$ and $\alpha_{\bot}$ respectively.
These are approximately twice of the errors obtained for the blind mock and taken a maximum over the two RSD models as given in Table~\ref{tab:sys-error}.
We would like to emphasis that these systematic errors are very conservative and are not a reflection of the limits of two RSD models but rather they reflect the limits of the tests performed in this work. In fact, the two RSD models (CLPT and TNS) are found to be unbiased at the precision of the results in this paper.

\begin{table}
	\centering
	\caption{Estimate of systematic and statistical errors in the three RSD parameters $f\sigma_8$, $\alpha_{\parallel}$ and $\alpha_{\bot}$. 
	The first three columns are from power spectrum measurements using the TNS model, whereas the second set of columns are from the correlation function measurements using the CLPT model. The sub-tables are for different ELG models, from top to bottom: SHOD, HMQ, SFHOD, and Blind models. 
	The quantity $\mu_{\rm sys}$ represent the mean bias, $\sigma_{\rm sys}$ represent the rms in the mean bias and $\sigma_{\rm stat}$ represent the statistical error.
	The values in the table are in units of $10^3$. We consider that there is a significant systematic bias when $\mu_{\rm sys}>2\sigma_{\rm stat}$. Note that all values reported in this table are absolute errors on the parameters and not percentage errors.}
	\setlength{\tabcolsep}{5pt}
	\setlength{\cmidrulekern}{0.45em}
	\begin{tabular}{l ccc | ccc} 
		\hline
		& \multicolumn{3}{c}{$P_{\ell}^{\rm TNS}$} & \multicolumn{3}{c}{$\xi_{\ell}^{\rm CLPT}$} \\
		\cmidrule(lr){2-4}\cmidrule(lr){5-7}
$\times 10^3$ & $f\sigma_8$ & $\alpha_{\parallel}$ & $\alpha_{\bot}$  
		 & $f\sigma_8$ & $\alpha_{\parallel}$ & $\alpha_{\bot}$ \\
		\hline
 {\bf SHOD} \\ 
 $\mu_{\rm sys}$  &  3.6  &  3.0  &  3.4  & 10.1  &  7.0  &  1.4  \\ 
$\sigma_{\rm sys}$  &  3.2  &  1.8  &  1.7  &  5.2  &  2.4  &  0.8  \\ 
$\sigma_{\rm stat}$  &  4.4  &  5.1  &  3.1  &  7.8  &  4.7  &  4.5  \\ 
 \hline {\bf HMQ} \\ 
 $\mu_{\rm sys}$  &  4.6  &  2.8  &  2.7  & 11.3  &  6.9  &  1.9  \\ 
$\sigma_{\rm sys}$  &  2.1  &  1.9  &  1.6  &  5.7  &  2.3  &  1.2  \\ 
$\sigma_{\rm stat}$  &  4.4  &  5.1  &  3.1  &  7.8  &  4.7  &  4.5  \\ 
 \hline {\bf SFHOD} \\ 
 $\mu_{\rm sys}$  &  4.7  &  4.7  &  3.2  &  8.1  &  8.3  &  3.2  \\ 
$\sigma_{\rm sys}$  &  3.7  &  3.7  &  2.2  &  4.7  &  4.1  &  2.2  \\ 
$\sigma_{\rm stat}$  &  5.8  &  6.0  &  3.7  &  9.3  &  5.8  &  4.9  \\ 
 \hline {\bf Blind} \\ 
 $\mu_{\rm sys}$  &  4.0  &  1.7  &  3.8  &  6.3  &  3.9  &  1.7  \\ 
$\sigma_{\rm sys}$  &  0.9  &  1.3  &  1.5  &  3.9  &  2.2  &  1.1  \\ 
$\sigma_{\rm stat}$  &  7.1  &  7.9  &  6.9  &  7.3  &  9.2  &  7.3  \\ 
 \hline
\hline
	\end{tabular}
	\label{tab:sys-error}
\end{table}

\section{Conclusions}
\label{sec:conclusion}
Galaxy redshift surveys \citep[e.g.][]{2dFGRS,BOSS,WiggleZ,6dFGRS,Vipers,gama2015,2016AJ....151...44D} measure three dimensional positions of millions of galaxies in redshift space. This allows us to measure the clustering of galaxies in redshift space in the late time Universe and, hence, to probe the cosmological growth at the epoch of the galaxy sample. Such measurement requires predicting the measured clustering in order to obtain constraints on the parameters of interest. But in principle this would require understanding galaxy physics to be able to predict the galaxy clustering at very high precision which is a very hard and highly non-linear problem. Therefore, the models often take a perturbative approach to solve the clustering of the dark matter and then perform another perturbative expansion of the galaxy formation process in terms of galaxy bias. Such solutions are expected to work very well on linear scales and for mass selected complete samples of galaxies.
When we start observing a wide varity of galaxies which might be highly influenced by their environment then such perturbative approach needs to be tested rigorously to avoid erroneously biased measurement of properties of the Universe. 

In this paper we focus on the eBOSS Emission Line Galaxy (i.e. star-forming galaxy) sample \citep{anand20}. eBOSS ELGs have relatively lower mass galaxies \citep{gp18,guo2019,2019arXiv191005095A} compared to luminous red galaxies studied in the past \citet{2017MNRAS.470.2617A}. The eBOSS ELG sample is analysed in Fourier space using the TNS model \citep{de-mattia20} and in configuration space using the CLPT model \citep{tamone20} for redshift space distortions. In this paper, we test these models, close to percent level precision, for the existence of any systematic bias due to theoretical approximations taken in the perturbative approach and simplistic ways to model the galaxy formation effects. This is the first tests being done at such high precision for these RSD models focusing on ELGs. This work should be considered as the first step towards testing of models for future surveys like \citep[DESI:][]{2016arXiv161100036D} and \citep[PFS:][]{PFS} which will be dominated by star-forming ELGs. Similar studies have been performed in companion papers focusing on the eBOSS QSOs sample \citet{smith20} and eBOSS LRG sample \citet{rossi20}.

We use high resolution N-body simulations, MultiDark and {\sc Outer Rim}, to obtain halo catalogs at the mean effective redshift ($z=0.86$) of the eBOSS ELGs sample. Such halo catalogue creates a fully non-linear realisation of the dark matter field which is the first essential ingredient of the Universe. We then populate the halo catalogues with a range of halo occupation distribution (HOD) models. We use three different parametrisations for the shape of the mean HOD of central galaxies. The first parametrization, SHOD, is the standard HOD which ignores existence of any galaxy quenching at the centre of massive haloes and is more appropriate for modelling magnitude or stellar mass selected samples \citep{zheng2005,White2011}. But we do allow the normalisation of the central occupation to be free to account for incompleteness of ELG in high mass dark matter haloes. 
The second HOD parametrization, HMQ, which encapsulates the quenching of the star formation in galaxies at the centre of massive haloes, and hence should provide a more realistic realisation of star-forming ELGs \citep{2019arXiv191005095A}. The third HOD parametrization, SFHOD, is based on the results for ELGs from a semi-analytical model of galaxy formation and evolution\citep{gp18,avila20}. In each of these HOD models we introduce parameters to account for other various baryonic effects that can affect the spatial distribution of satellite galaxies, their dynamical properties, including infalling velocities, assembly bias, off centring in the location of central galaxies and deviations in large-scale velocities.

We first create a set of non-blind mock catalogues, for which all the parameters of the mocks were available to the teams analysing them. We then analyse these non-blind mock based on MultiDark and {\sc Outer Rim} using the TNS model in Fourier space and the CLPT model in configuration space. For the mocks based on the MultiDark simulation, illustrative power spectrum are shown in Figure ~\ref{fig:MDPL2-nonblind-pk}, the result of the RSD analysis is shown in Figure ~\ref{fig:MDPL2-RSDfit} and the parameters constraints are given in Table~\ref{tab:MDPL2-RSDresults}. We note that Figure 2 highlights that the impact of galaxy physics on the galaxy power spectrum can be up to 10\% by $k \approx 0.2$ (scales analysed in this paper). The MultiDark mocks have a volume of $1 ({\rm Gpc/h})^3$ and hence have statistical errors of ~8\% , 3\% and 2\% in $f\sigma_8$, $\alpha_{\parallel}$ and $\alpha_{\bot}$, respectively. We do not detect any significant bias when analysing the MultiDark mocks with either of the RSD models. But this is a weak statement given the statistical uncertainty of these mock due to their small volume. For the non-blind mocks based on the {\sc Outer Rim} simulation, the results of the RSD analysis are shown in Figure ~\ref{fig:OR-RSDfit} and the parameters constraints are given in Table~\ref{tab:OR-RSDresults} and ~\ref{tab:OR-SFHOD-RSDresults}. 
The statistical uncertainties in the {\sc Outer Rim} mocks with volume of $27 ({\rm Gpc/h})^3$ are ~1-2\%, 0.5-0.6\% and 0.3-0.5\% in $f\sigma_8$, $\alpha_{\parallel}$ and $\alpha_{\bot}$, respectively. We note that for {\sc Outer Rim} non-blind mocks we again do not detect any statistical significant systematic bias at the level of statistical uncertainty in these parameters despite the wide range of ELG models used.

We have analysed a wide variety of models, with a range of kinematical degrees of freedom for satellite galaxies, assembly bias and various forms of mass incompleteness. Nevertheless, we do not span the complete parameter space of the ELG connection to the dark matter haloes and cosmic web. For example we do not consider any model that correlates the ELG occupation with the tidal environment of dark matter haloes. \cite{2019arXiv191005095A} recently showed that the ELGs slightly prefers to populate the haloes in the low density filaments compared to the prediction of HMQ model used in this paper.
But any such tidal correlation in observational data have only been detected at low significance and hence are expected to be small.  Therefore, at the level of our precision we suggest that our models spans wide enough parameter space of ELG population such that we can be confident about the robustness of RSD models.

We finally create a set of blind mocks. Our focus has been to study the biases in the $f\sigma_8$ coming from theoretical approximations in the RSD models. Therefore, our mocks are blind only in the $f\sigma_8$ measurements and all other information was known to the analysis teams. We show the results of the RSD analyses from blind mocks in Figure ~\ref{fig:OR-RSD-blind} and the constraints are shown in Table~\ref{tab:OR-RSDresults-blind}. We also show the comparison of mock measurements and best-fit models for the blind mocks in Figure ~\ref{fig:pk-xi-blind-wbfit}. Based on these figures and tables we conclude that the TNS model in Fourier space and the CLPT model in configuration space can describe the blind mock catalogues remarkably well, obtaining unbiased measurement of $f\sigma_8$.

We present the systematic error from all the mocks in \S ~\ref{sec:sys}. Figure ~\ref{fig:sys-err} presents the systematic errors and Table~\ref{tab:sys-error} lists their values for both the RSD models, comparing them to statistical errors for the different categories of mocks. We conclude, through these series of detailed analysis of mocks with versatile galaxy physics models, that the TNS model in Fourier space and the CLPT model in configuration space provide an unbiased measurement of redshift space distortions within the statistical error of our mocks. Therefore, taking a conservative choice, we suggest using twice the statistical error obtained for the blind mocks as the theoretical systematic for these model unless a more precise test is performed. 
For both RSD models (i.e. TNS and CLPT), we propose the common theoretical systematic errors of 3.3\%, 1.8\% and 1.5\% in $f\sigma_8$, $\alpha_{\parallel}$ and $\alpha_{\bot}$, respectively.
The theoretical systematic errors proposed here are an order of magnitude smaller than the statistical error for eBOSS ELG sample \citet{de-mattia20,tamone20} and hence are negligible for the purpose of the current eBOSS ELG analysis. We emphasise that redshift space distortions of incomplete galaxy samples such as ELGs can be modelled with TNS ($k_{\rm max}=0.2\hMpc$) and CLPT ($s_{\rm min}=32 \, h^{-1}\mathrm{Mpc} $) without any systematic biases to a few percents level. 

The upcoming DESI survey \citep{2016arXiv161100036D} will have an effective volume of $~20 ({\rm Gpc/h})^3$ for the ELG sample. This will result in statistical errors smaller than the systematic errors proposed in this paper. Hence, systematic errors can have a significant contribution to the total error budget for DESI ELGs. Therefore, one must perform a similar analysis with much smaller uncertainty and hence much bigger volume of simulations in order to avoid adding significant uncertainty from theoretical systematic to the total error budget.

\section{Data Availability}
All of the observational datasets used in this paper are available through the SDSS website \url{https://data.sdss.org/sas/dr16/eboss/}. The codes used in this analysis along with instructions are available on \url{https://www.roe.ac.uk/~salam/MTHOD/}. The outer rim halo catalogues are available on \url{https://cosmology.alcf.anl.gov/}.

\section*{Acknowledgements}
SA and JAP are supported by the European Research Council
through the COSFORM Research Grant (\#670193). VGP acknowledges support from the European Union's Horizon 2020 research and innovation programme (ERC grant \#769130) and by the Atracci\'{o}n de Talento Contract no. 2019-T1/TIC-12702 granted by the Comunidad de Madrid in Spain. SH and KH acknowledge support under the U.S. Department of Energy contract W-7405-ENG-36 at Argonne National Laboratory. GR acknowledges support from the National Research Foundation of Korea (NRF) through Grants No. 2017R1E1A1A01077508 and No. 2020R1A2C1005655 funded by the Korean Ministry of Education, Science and Technology (MoEST). RP and SE acknowledge support from the French National Research Agency under Grants No. ANR-16-CE31-0021 (ANR eBOSS), No. ANR-11-LABX-0060 (OCEVU LABEX) and No. ANR-11-IDEX-0001-02 (A*MIDEX project).

Funding for the Sloan Digital Sky Survey IV has been provided by the Alfred P. Sloan Foundation, the U.S. Department of Energy Office of Science, and the Participating Institutions. SDSS-IV acknowledges
support and resources from the Center for High-Performance Computing at
the University of Utah. The SDSS web site is www.sdss.org.

SDSS-IV is managed by the Astrophysical Research Consortium for the 
Participating Institutions of the SDSS Collaboration including the 
Brazilian Participation Group, the Carnegie Institution for Science, 
Carnegie Mellon University, the Chilean Participation Group, the French Participation Group, Harvard-Smithsonian Center for Astrophysics, 
Instituto de Astrof\'isica de Canarias, The Johns Hopkins University, Kavli Institute for the Physics and Mathematics of the Universe (IPMU) / 
University of Tokyo, the Korean Participation Group, Lawrence Berkeley National Laboratory, 
Leibniz Institut f\"ur Astrophysik Potsdam (AIP),  
Max-Planck-Institut f\"ur Astronomie (MPIA Heidelberg), 
Max-Planck-Institut f\"ur Astrophysik (MPA Garching), 
Max-Planck-Institut f\"ur Extraterrestrische Physik (MPE), 
National Astronomical Observatories of China, New Mexico State University, 
New York University, University of Notre Dame, 
Observat\'ario Nacional / MCTI, The Ohio State University, 
Pennsylvania State University, Shanghai Astronomical Observatory, 
United Kingdom Participation Group,
Universidad Nacional Aut\'onoma de M\'exico, University of Arizona, 
University of Colorado Boulder, University of Oxford, University of Portsmouth, 
University of Utah, University of Virginia, University of Washington, University of Wisconsin, 
Vanderbilt University, and Yale University.

In addition, this research relied on resources provided to the eBOSS
Collaboration by the National Energy Research Scientific Computing
Center (NERSC).  NERSC is a U.S. Department of Energy Office of Science
User Facility operated under Contract No. DE-AC02-05CH11231.

The CosmoSim database used in this paper is a service by the Leibniz-Institute for Astrophysics Potsdam (AIP).
The MultiDark database was developed in cooperation with the Spanish MultiDark Consolider Project CSD2009-00064.

The authors gratefully acknowledge the Gauss Centre for Supercomputing e.V. (www.gauss-centre.eu) and the Partnership for Advanced Supercomputing in Europe (PRACE, www.prace-ri.eu) for funding the MultiDark simulation project by providing computing time on the GCS Supercomputer SuperMUC at Leibniz Supercomputing Centre (LRZ, www.lrz.de).




\bibliographystyle{mnras}
\bibliography{Master_shadab} 







\bsp	
\label{lastpage}
\end{document}